\documentclass[journal]{IEEEtran}
\usepackage[utf8]{inputenc}
\usepackage{amsmath}
\usepackage{float}
\floatstyle{plaintop}
\restylefloat{table}
\usepackage{amsthm}
\usepackage{amssymb}
\usepackage{graphicx}
\usepackage{cite}
\usepackage{tablefootnote}
\usepackage{footnote}
\usepackage{float}
\usepackage{comment}
\usepackage{epstopdf}
\usepackage{float}

\usepackage{algorithmicx}
\usepackage{amsfonts}
\usepackage{subcaption}
\usepackage{tikz}

\usepackage[noend]{algpseudocode}
\usepackage[ruled,norelsize,linesnumbered]{algorithm2e}
\usepackage{multicol}
\usepackage{circledsteps}

\makeatletter
\newcommand{\removelatexerror}{\let\@latex@error\@gobble}
\g@addto@macro{\@algocf@init}{\SetKwInOut{Parameter}{Parameters}}
\makeatother

\algdef{SE}[DOWHILE]{Do}{doWhile}{\algorithmicdo}[1]{\algorithmicwhile\ #1}%
\makeatletter
\def\BState{\State\hskip-\ALG@thistlm}

\newlength\myindent
\title{\Large\bf Simultaneous Suspension Control and Energy Harvesting through Novel Design and Control of a New Nonlinear Energy Harvesting Shock Absorber  }
\author{ Mohammad R. Hajidavalloo$^{1}$, Joel Cosner$^{1}$, Zhaojian Li$^{*1}$, Wei-Che Tai$^{1}$, Ziyou Song$^{2}$
\thanks{\quad  Mohammad Hajidavalloo, Joel Cosner, Zhaojian Li and Wei-Che Tai are with the Department of Mechanical Engineering, Michigan State University, East Lansing, MI 48824, USA.
        Email: {\tt\small \{hajidava,cosnerjo, lizhaoj1,taiweich\}
        @egr.msu.edu}.}
\thanks{\quad  Ziyou Song is with the Department of Mechanical Engineering, National University of Singapore, Singapore, 117575 Singapore.
        Email: {\tt\small ziyou@nus.edu.sg}.}
\thanks{$*$ Zhaojian Li is the corresponding author.}}

\IEEEoverridecommandlockouts
\begin{document}

\maketitle
\begin{abstract}
Simultaneous vibration control and energy harvesting of vehicle suspensions have attracted significant research attention over the past decades. However, existing energy harvesting shock absorbers (EHSAs) compromise suspension performance for high-efficiency energy harvesting and being only responsive to narrow-bandwidth vibrations. In this paper, {we propose a new ball-screw-based EHSA design -- inerter pendulum vibration absorber (IPVA)} -- that integrates an electromagnetic rotary EHSA with a nonlinear pendulum vibration absorber.{We show that this design simultaneously improves ride comfort and energy harvesting efficiency by exploiting the  nonlinear effects of pendulum inertia.} To further improve the performance, {we develop a novel stochastic linearization model predictive control (SL-MPC) approach} in which we employ stochastic linearization to approximate the nonlinear dynamics of EHSA that has superior accuracy compared to standard linearization. In particular, we develop a new stochastic linearization method with guaranteed stabilizability, which is a prerequisite for control designs. This leads to an MPC problem that is much more computationally efficient than the nonlinear MPC counterpart with no major performance degradation. {Also, the effect of different road preview configurations on control performance is investigated, which is shown to have a significant impact on the control performance.} \textcolor{black}{ Extensive simulations are performed }to show the superiority of the proposed new nonlinear EHSA and to demonstrate the efficacy of the proposed SL-MPC. 

\end{abstract}

\section{Introduction}
Traditional vehicle suspensions use hydraulic dampers to dissipate undesired vibration energy into heat waste, thereby improving ride comfort. There exists a great potential for harvesting this wasted energy; it is estimated that somewhere between 100W to 10kW of power per vehicle can be harvested for an average trip  \cite{abdelkareem2018,chen2020mpcbased}. The growth in hybrid and electric vehicles have further increased the potential impact through smart utilization and management of this harvested energy \cite{shi2016}. Therefore, extensive and increasing research efforts over the past three decades have focused on developing energy harvesting shock absorbers (EHSAs) -- devices that convert vehicle suspension vibrations into useful electricity. Traditionally, EHSAs use electromagnetic dampers or magnetorheological dampers to recycle this heat waste into useful electricity. Rotary electromagnetic dampers have become popular because of their high conversion efficiency and quick responsiveness \cite{abdelkareem2018}. Thanks to their quick responsiveness, electromagnetic dampers are integrated with power electronic circuits to perform damping force control in real time \cite{hsieh2015bidirectional,xie2018damping} or used as actuators to deliver active force to improve road handling and ride comfort \cite{gysen2011efficiency}.

\textcolor{black}{To drive the rotary electromagnetic (EM) damper, motion conversion mechanisms, such as rack-pinion \cite{li2012electromagnetic} and ball-screw \cite{hoo2013investigation}, are required to convert the linear suspension vibrations into angular motion. This operation principle, although being straightforward, has a critical drawback; that is, they require large suspension vibrations to achieve high-efficiency energy harvesting, thereby compromising suspension performance for energy recovery. This drawback has been widely recognized in the literature. Through numerical simulations on a quarter car, Abdelkareem et al. \cite{abdelkareem20182} concluded that ride comfort and the harvestable power cannot be optimized at the same time. Through numerical simulations on a quarter car with a traditional EHSA, Casavola et al. \cite{casavola2018} showed that a trade-off always exists between road handling and the energy harvesting performance. Huang et al. \cite{huang2015} considered a traditional ball-screw-based EHSA in a quarter car and discovered that ride comfort and the harvested power are conflicting objectives. Guo et al. \cite{guo2016} and Li and Zuo \cite{li2017} considered a traditional rack-pinion-based EHSA in a quarter car and showed that ride comfort and the harvested power cannot be optimized together.}

Therefore, there is a pressing need for fundamentally new EHSA designs to fully realize the potential benefits of harvesting vehicle vibration energy while simultaneously achieving great suspension performance. \textcolor{black}{Recently, Gupta and Tai proposed a nonlinear rack-pinion-based EM damper, known as inerter pendulum vibration absorber (IPVA) \cite{gupta2020broadband, 9658755}. The IPVA consists of a planetary gear set that integrates a rack-pinion-based EM damper and a nonlinear pendulum vibration absorber. It was shown that the nonlinear inertial effects of the pendulum increased the harvested power and energy harvesting bandwidth when subject to harmonic excitation. Although showing promising results, the rack-pinion mechanism is too bulky to fit in a typical vehicle suspension system. Later, Cosner and Tai \cite{cosner2021vibration} proposed to use a ball-screw in lieu of rack-pinion, which is more compact and suitable for vehicle suspension systems. However, their design was not able to integrate with a rotary EM damper for electricity generation, and only capable of vibration suppression. Nevertheless, they showed that the pendulum absorbed the vibration energy of a suspended platform when subject to white noise excitation, thereby holding promise to achieve energy harvesting and vibration suppression at the same time.  In this paper, we propose a new design that integrates the pendulum vibration absorber and a ball-screw-based EM damper. This new ball-screw-based IPVA is integrated with a quarter-car suspension model where we optimize the design parameters to achieve a better tradeoff between the suspension performance (ride comfort) and energy harvesting efficiency of the system when subject to stochastic road excitation. }

To further improve the energy harvesting efficiency while
maintaining good ride comfort, model predictive
control (MPC) is applied to the IPVA-integrated quarter-car suspension model. MPC is a popular tool for solving constrained optimal control problems with the advantage of online implementation relative to methods such as dynamic programming \cite{rawlings2017model, Me1, Me2, 9296781}. Since the IPVA dynamics and the objective function (as defined in the sequel) are nonlinear, a nonlinear MPC (NMPC) can be exploited to solve this problem. However, the NMPC problem is computationally expensive and is difficult for real-time implementations, especially for suspension systems that require very high control frequency. To address this issue, we propose a new MPC framework  by exploiting an approximated linear dynamics using the technique of stochastic linearization \cite{roberts2003random,kozin1988method}. The new MPC framework with the stochastically linearized dynamics has comparable control performance with NMPC while requiring significantly less computation power. Moreover, the MPC performance is also investigated with and without the usage of perfect road preview, which can be obtained through recent road information estimation techniques using a single or multiple of vehicles \cite{li33,li34}. We show that the usage of road profile preview can greatly improve the performance.

\textcolor{black}{The contributions of this paper include the following.
First, we integrate IPVA into automotive suspension systems and optimize the system designs that offer improved ride comfort and energy harvesting efficiency at the same time when compared to the traditional EHSA.  
Second, we develop a novel stochastic-linearization MPC (SL-MPC) framework by exploiting a stochastically-linearized dynamics based on the nonlinear equation of motion (EOM) of the IPVA. To the best of the authors{'} knowledge, this is the first time that stochastic linearization is exploited in MPC designs to deal with nonlinear dynamical systems excited by random disturbance signals.
Third, we investigate the usage of online estimated road information into the prediction horizon, which we show is able to enhance the performance. Last but not least, extensive simulations are performed to demonstrate the efficacy of the proposed framework.} \textcolor{black}{It should be noted that although Chen et al. \cite{chen2020mpcbased} proposed a nonlinear EHSA that also utilized inertial nonlinearity, their nonlinear EHSA was only efficacious around a resonance peak; that is, it is narrow-banded. Furthermore, their numerical study showed that the maximum energy harvesting efficiency and worst ride comfort occurred at the same frequency. In other words, energy harvesting and suspension performance are still conflicting objectives in their design.}

The rest of this paper is organized as follows. In Section~II,
the design and modeling of the IPVA-integrated suspension system are introduced. Section~III describes the NMPC and SL-MPC designs. Simulations and performance evaluations are presented
in Section~IV while Section~V concludes the paper.

\section{Energy Harvesting System Description}
\textcolor{black}{In this section, we introduce a novel nonlinear EHSA design,  along with a  linear benchmark model (subsection II-A). The governing equations of motion (EOM) for both systems are derived (subsection II-B). The optimal parameter designs for both systems are also discussed (subsection II-c).}
 \subsection{System structure}\label{sec:sys_struct}
\begin{figure}[ht!]
    \centering
    \includegraphics[width=0.45\linewidth] {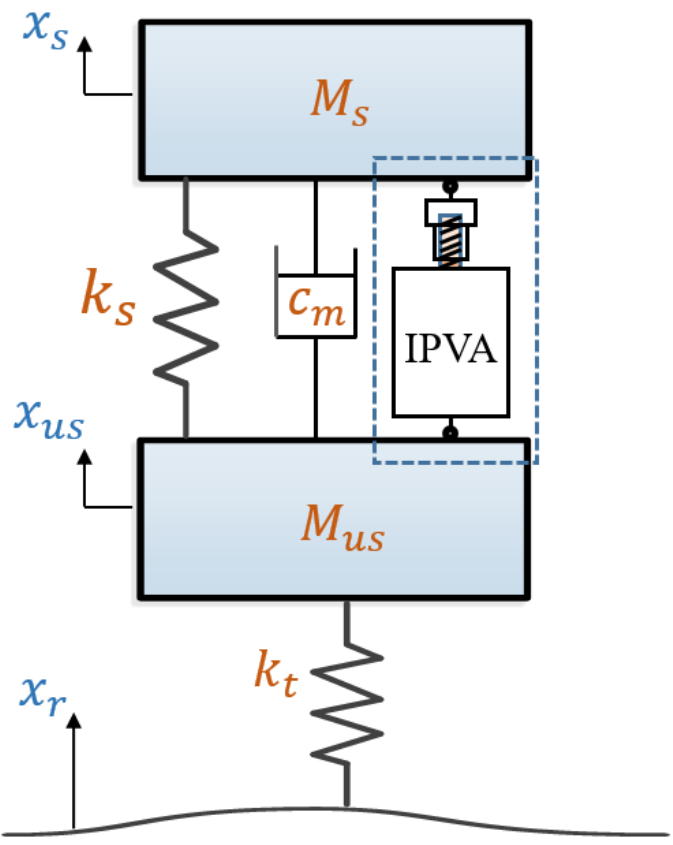}
    \caption{Schematics of an IPVA-integrated quarter-car model.}
    \label{fig:quartcar_IPVA}
\end{figure}

\begin{figure*}
     \centering
     \begin{subfigure}[b]{0.3\textwidth}
         \centering
         \includegraphics[width=\textwidth]{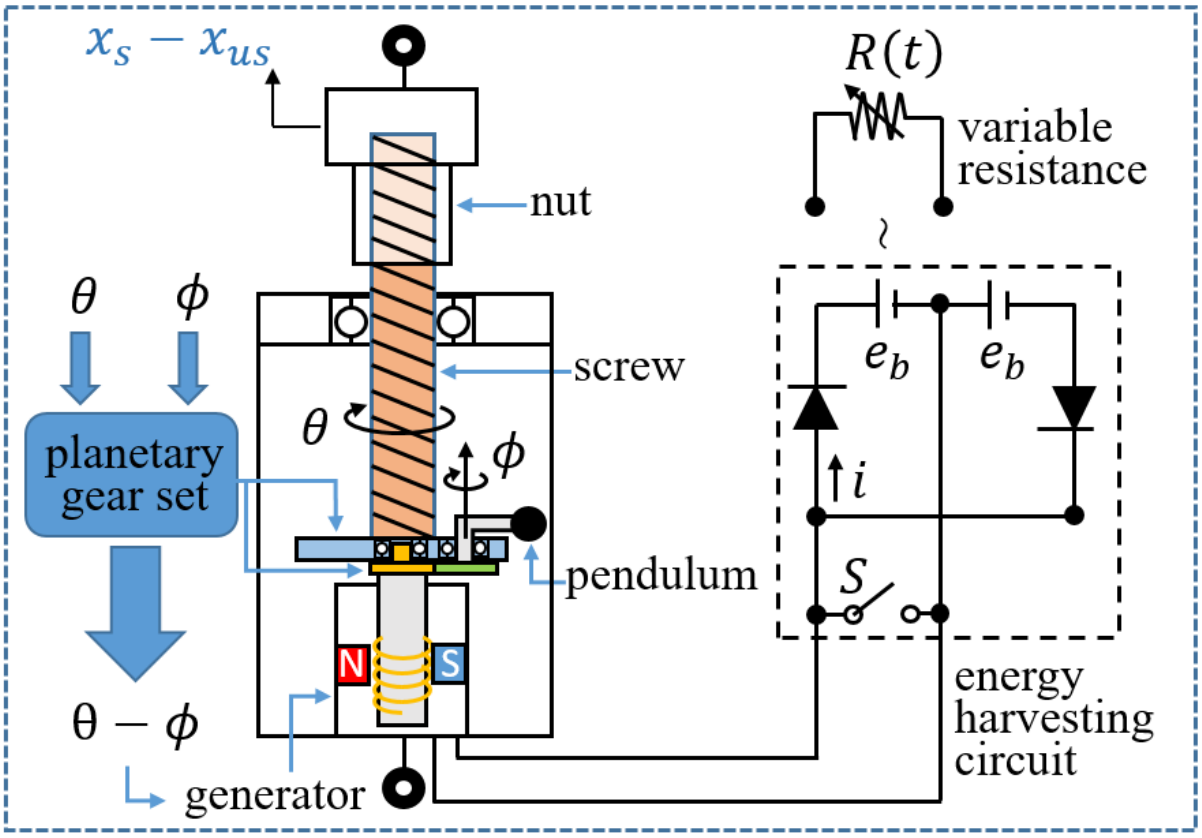}
         \caption{Conceptual design of semi-active inerter pendulum vibration absorber.}
         \label{fig:IPVA}
     \end{subfigure}
     \hfill
     \begin{subfigure}[b]{0.3\textwidth}
         \centering
         \includegraphics[width=\textwidth]{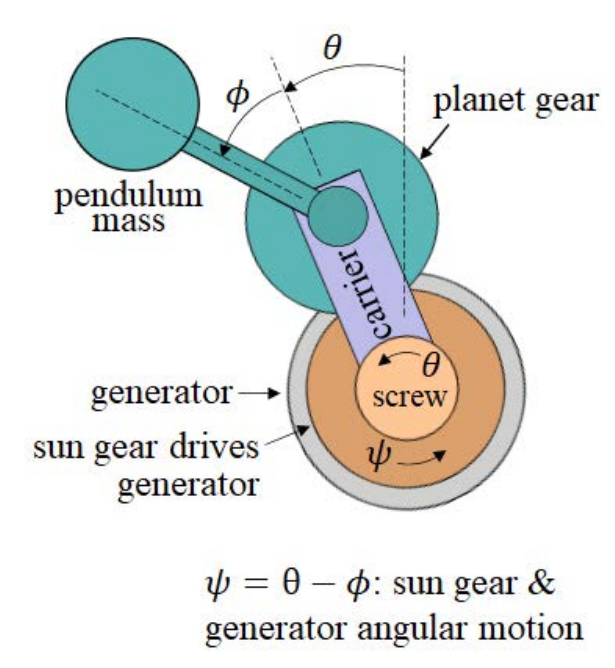}
         \caption{Top view of pendulum and gear arrangement}
         \label{fig:IPVA_pend}
     \end{subfigure}
     \hfill
     \begin{subfigure}[b]{0.3\textwidth}
         \centering
         \includegraphics[width=\textwidth]{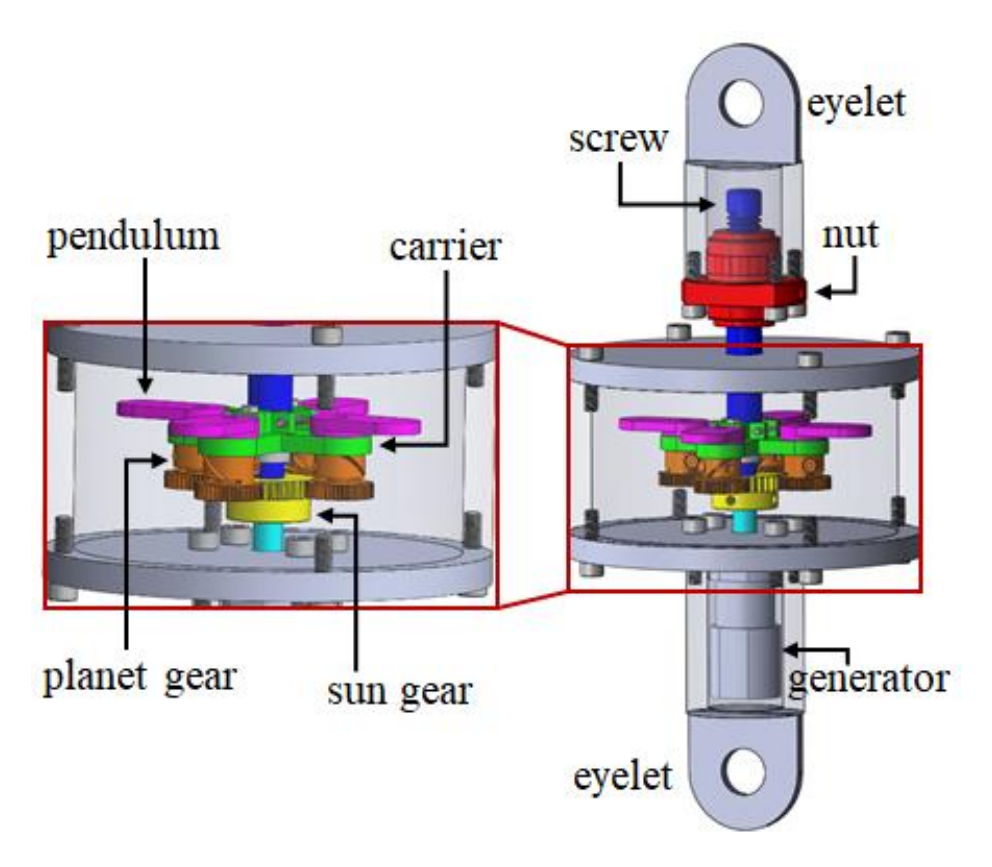}
         \caption{Realization of a possible prototype.}
         \label{fig:IPVA_CAD}
     \end{subfigure}
        \caption{\textcolor{black}{Schematic for the IPVA.}}
        \label{fig:IPVA_3figs}
\end{figure*}

The IPVA-integrated quarter-car suspension model is shown in Fig.~\ref{fig:quartcar_IPVA}, where the mass of a quarter-car body and the unsprung mass (i.e., wheel axle) of a quarter car are represented by $M_s$ and $M_{us}$, respectively. The interaction between the unsprung mass $M_{us}$ and the ground is characterized by a spring of stiffness $k_t$ (tire stiffness) with its displacement from the equilibrium given by $x_{us}$.  The quarter-car model is excited by the road profile/disturbance $x_r$.  The sprung mass and unsprung mass are connected with a spring of stiffness $k_s$ (suspension stiffness), a viscous damper with damping coefficient $c_m$, and the IPVA. \textcolor{black}{Note that the mechanical damping $c_m$ is introduced to account for mechanical energy loss due to the ball-screw and generator gears.} As shown in Fig.~\ref{fig:IPVA}, the IPVA consists of a ball-screw system with a lead $L$ connected between $M_{us}$ and $M_s$. It converts the linear oscillations between $M_{us}$ and $M_s$ to the rotation of the screw. The carrier fixed to the screw houses a pendulum of mass $m$ and radius $r$ at a distance of $R_p$ from the screw's axis of rotation. A sun gear is free to rotate with respect to the carrier about the same axis of rotation as the screw and drives the generator. A planet gear fixed to the pendulum rotates and revolves on the sun gear. The housing of the generator is fixed to the sprung mass. Considering the gear ratio $g_p$ between the sun and planet gear to be 1, we have $\psi = \theta -\phi$ \color{black}as shown in Fig.~\ref{fig:IPVA_pend}, $\psi$, $\theta$, and $\phi$ are the angular displacement of the generator rotor, screw, and pendulum, respectively. \color{black}The generator is connected with an electricity storage $e_b$ via a pulse width modulated (PWM) step-up chopper to modulate the duty cycle of the generator by switch $S$, equivalent to a variable resistor $R(t)$ \cite{okada2002}. The corresponding electrical damping coefficient in the generator is denoted by $c_e(t)=M\kappa_t\kappa_e/R(t)$ \cite{zuo2013energy}, where $M$ is a motion transmission factor that is related to the ball screw lead and gear ratio of the generator, and $\kappa_t$ and $\kappa_e$ are the torque and voltage constant of the generator, respectively. \textcolor{black}{Note that the generator's inductance is neglected because the impedance of the inductance is small compared with the resistance considering that the vibration induced by road irregularities is usually in the frequency range of 1–10 Hz \cite{li2012electromagnetic,hsieh2015bidirectional}.} This model is a three degrees of freedom (DOF) system, with the degrees of freedom being the angular displacements of the pendulum ($\phi$) and the screw ($\theta$), and the displacement of the unsprung mass $x_{us}$. \textcolor{black}{Note that the sprung mass displacement $x_{s}$ is related to $\theta$ and $x_{us}$ via $x_{s}-x_{us}=R\theta$, where $R=L/2\pi$.} \color{black}Finally, a 3D model for a potential prototype for the proposed IPVA is shown in Fig.~\ref{fig:IPVA_CAD}. While the prototype shown consists of four pendulums, the pendulums move synchronously due to the planetary gear system. Hence this prototype is equivalent to a single pendulum system with a quadrupled pendulum mass $m$. \color{black}

\subsection{Equations of motion of the EHSA}
We next use the Lagrangian method to derive the EOM for the IPVA-integrated quarter-car system shown in Fig.~\ref{fig:quartcar_IPVA}. \textcolor{black}{The kinematic relation $x_s - x_{us} =R\theta$ relates the suspension deflection $x_s-x_{us}$ to the angular displacement $\theta$ with $R=L/2\pi$, from which it follows that the total kinetic energy of the system is:}
\begin{equation}\begin{aligned}
T &= {T}_{M_{us}}+T_M+T_c + T_p + T_r \\ 
&={1\over 2} M_{us}\left( \dot{x}_{us}\right)^2+{1\over 2} M_s\left(R\dot{\theta} + \dot{x}_{us}\right)^2+{1\over 2} J \dot{\theta}^2\\&+ {1\over 2}m\left(R_p^2\dot{\theta}^2 + r^2\left(\dot{\theta}+\dot{\phi}\right)^2+2R_pr\cos\left(\phi\right)\dot{\theta}\left(\dot{\theta} + \dot{\phi}\right)\right) \\
&+  {1\over 2}J_p\left(\dot{\theta} + \dot{\phi}\right)^2 + {1\over 2} J_r  (\dot{\phi}-\dot{\theta})^2,
\label{eq:kinetic}\end{aligned}
\end{equation}
where $R_p$ and $r$ are the distance between the pendulum pivot point and center of the carrier (i.e., half carrier length), and the length of the pendulum, respectively; the parameters $J$, $J_r$, and $J_p$ represent
the principal moment of inertia (w.r.t. primary rotational axis) of the carrier, generator rotor, and pendulum, respectively. Note that the moment of inertia of the gears and screw are assumed to be negligible. The deformation of the springs contribute to the potential energy, which can be obtained as:
\begin{gather}
V = {1\over 2}k_sR^2\theta^2+{1\over 2}k_{t}(x_{us}-x_r)^2 + {1\over 2}k_p \phi^2.
\end{gather}

The last term is related to the torsion spring attached to the pendulum which has a very low stiffness for making sure the stochastic linearized model (discussed in Section~III-B) is stable. \color{black}We use the concept of virtual work to include the non-conservative forces in the system. Since the virtual angular displacement of the rotor is $\delta\left( \psi\right)=\delta\left( \theta-\phi\right)$, the virtual work on the rotor by the damping torque due to energy harvesting is $\delta W_r = -c_e \left(\dot{\theta}-\dot{\phi}\right)\cdot\delta\left( \theta-\phi\right)$. Here $\delta$ represents the variational operator. For the mechanical damper, the virtual displacement of the mass is $R\cdot\delta \theta$, so the virtual work is given by $\delta W_m = -c_mR^2 \dot{\theta}\cdot\delta \theta$. \color{black}Therefore, the total virtual work due to non-conservative forces is:
\begin{equation}
\begin{aligned}
\centering
  	\delta W_{nc} &= \delta W_r + \delta W_m =-c_e (\dot{\phi}-\dot{\theta})\cdot\delta\phi\\&+ c_e (\dot{\phi}-\dot{\theta})\cdot\delta\theta-c_m R^2\dot{\theta} \cdot\delta \theta.
\end{aligned}
\end{equation}

As a result, the viscous damping forces included in the EOM for $\theta$ and $\phi$ are:
\begin{align*}
	&\theta: \ -c_m R^2\dot{\theta} +c_e (\dot{\phi}-\dot{\theta}),\\
	&\phi: \ -c_e (\dot{\phi}-\dot{\theta}).
\end{align*}
The Lagrange equations are derived as,
\begin{equation}
\frac{d}{d t} (\frac{\partial L}{\partial \dot{q}_{i}})-\frac{\partial L}{\partial q_{i}}=Q_{i},
\label{Eq:lagrange}
\end{equation}
where $L=T-V$ is the Lagrangian, $q_{1}=\theta$, $q_{2}=\phi$ and $q_{3}=x_{us}$ are the degrees of freedom and $Q_{i}$ is the generalized force for degree of freedom $q_i$. After substituting the corresponding terms into Eq. \ref{Eq:lagrange}, the EOM is obtained as:
\begin{equation}\label{eq:EOM}
    \begin{aligned}
  & G_{22}\ddot{\theta }
  +G_{24}\ddot{\phi }+{{c}_{m}}{{R}^{2}}\dot{\theta }-{{c}_{e}}(\dot{\phi }-\dot{\theta })
  +k{{R}^{2}}\theta\\
  &-2m{{R}_{p}}r\dot{\phi }\dot{\theta }\sin (\phi )-m{{R}_{p}}r\sin (\phi ){{{\dot{\phi }}}^{2}}+R{{M}_{s}}{{{\ddot{x}}}_{us}}=0,
  \end{aligned}
\end{equation}
\begin{equation}
    \begin{aligned}
  & G_{44}\ddot{\phi }+G_{42}\ddot{\theta }
 +{{c}_{e}}(\dot{\phi }-\dot{\theta }) +k_p \phi + {{R}_{p}}r\sin (\phi ){{{\dot{\theta }}}^{2}}=0 ,
\end{aligned}
\end{equation}
\begin{equation}\label{eq:EOM3}
    \begin{aligned}
    & ({{M}_{s}}+{{M}_{us}}){{{\ddot{x}}}_{us}}+{{M}_{s}}R\ddot{\theta }+{{k}_{t}}({{x}_{us}}-{{x}_{r}})=0,
  \end{aligned}
\end{equation}
  
where 
\[{{G}_{22}}={{M}_{s}}{{R}^{2}}+J+m{{R}_{p}}^{2}+m{{r}^{2}}+2m{{R}_{p}}r\cos (\phi )+{{J}_{p}}+{{J}_{r}},\]
\[{{G}_{24}=G_{42}}=m{{r}^{2}}+m{{R}_{p}}r\cos (\phi )+{{J}_{p}}-{{J}_{r}},\]
\[{{G}_{44}}=m{{r}^{2}}+{{J}_{p}}+{{J}_{r}}.\]

For simplicity, the pendulum mass is represented by a point mass and the moment of inertia of the carrier is assumed to be small, resulting in negligible $J_p$ and $J$.

\begin{table}[h!]
    \centering
    \scalebox{.7}{
    \begin{tabular}{|l|l|}
    \hline
       Parameters & Physical meanings \\ 
          \hline
       $M_s$, $M_{us}$, $m$ & mass of sprung, unsprung, and pendulum\\\hline
       $L$ & ball-screw lead\\\hline
       $R=L/(2\pi)$ & characteristic length of ball-screw\\\hline
       $R_p$ & distance between the pendulum pivot point and  center  of  the  carrier\\\hline
       $r$ & pendulum length\\\hline
       $J$, $J_r$, $J_p$ & principal moment of inertia of carrier, generator rotor, and pendulum\\\hline
       $k_s$, $k_t$, $k_p$ & stiffness of suspension, tire and pendulum torsion spring
       \\\hline
    \end{tabular}}
    \caption{\textcolor{black}{Parameters of IPVA system.}}

\end{table}

Note that for a traditional linear EM damper, the pendulum does not exist and consequently  $\phi$, $\dot{\phi}$, and all nonlinear terms will vanish. The non-conservative force for $\theta$ remains and the EOM for the linear benchmark is thus:
 \begin{equation}\label{lb1}
     \left({{M}_{s}}{{R}^{2}}+J_r\right)\ddot{\theta }+({{c}_{m}}{{R}^{2}}+{{c}_{e}})\dot{\theta }+k_s{{R}^{2}}\theta +RM_s{{\ddot{x}}_{us}}=0,
 \end{equation}
\begin{equation}\label{lb2}
\begin{aligned}
   (M_{us}+{{M}_{s}}){{\ddot{x}}_{us}}+M_{s}R\ddot{\theta}+{{k}_{t}}({{x}_{us}}-{{x}_{r}})=0.
\end{aligned}
\end{equation}

\subsection{Optimal design of IPVA and linear benchmark}
The performance of the IPVA depends on appropriate choices of design parameters, including   $R_p$, $r$, and $c_e$. Towards that end, we define the following dimensionless variables: 
\[\eta =\frac{r}{{{R}_{p}}},\text{  }{{\mu }_{r}}=\frac{mR_{p}^{2}}{M_s{{R}^{2}}},\text{  }{{\xi }_{e}}=\frac{{{c}_{e}}}{2{{\omega }_{0}}M_s{{R}^{2}}}.\]
The maximum of electrical damping coefficient $c_e$ depends on the internal resistance of generator. Considering the maximum electrical damping coefficient in ref.~\cite{abdelkareem20182}, it is assumed that $c_e<7.2\, N\cdot s\cdot m$, resulting in $\xi_e<1$. Furthermore, it is assumed that $\eta<0.9$ such that the pendulum length is smaller than the carrier radius for compactness and $\mu_r<0.2$ such that the pendulum mass $m<2.5$ kg for a reasonable weight. Then a reasonable choice of design parameters should satisfy the following constraints:
\begin{equation}\label{unitless}
   0.5<\eta <0.9,\quad 
 0.05<{{\mu }_{r}}<0.2{,\quad {\xi }_{e}}<1. \\ 
\end{equation}
\textcolor{black}{Note that the suspension spring stiffness is considered as given and not included in the optimization. The main reason is that we envision to fit our EHSA into existing suspension systems with given spring stiffness, according to which one can determine an optimal design and retrofit it to the suspension. Huang et al. \cite{huang2015} also fixed the suspension spring stiffness when determining local optimal designs of a traditional ball-screw-based EHSA.}
To get the optimal values for design parameters, a constrained vector objective optimization problem is defined with the variables $R_p$, $r$ and $c_e$, i.e.,
\begin{equation}\label{eq:optnon}
   \begin{matrix}
  \underset{{{c}_{e}},{{R}_{p}},r}{\mathop{\min .}}\,\text{    }\left[ \begin{matrix}
   \frac{1}{N}\sum\limits_{t=0}^{N}{-{{c}_{e}}{{(\dot{\phi }(t)-\dot{\theta }(t))}^{2}}\text{,}} & \sum\limits_{t=0}^{N}{\sqrt{\frac{1}{N}(\ddot{x}_s)^2}}  \\
\end{matrix} \right]^T \\ 
  \text{s}\text{.t}\text{.   }(\ref{eq:EOM})-(\ref{eq:EOM3}),\text{ }(\ref{unitless}).\text{  }
\end{matrix}
\end{equation}
Here the first element of the objective function vector represents the average harvested power and the second element represents the RMS value of the sprung mass acceleration, which is used to characterize the ride comfort.

\begin{figure}[!h]
    \centering
    \includegraphics[width=0.9\linewidth]{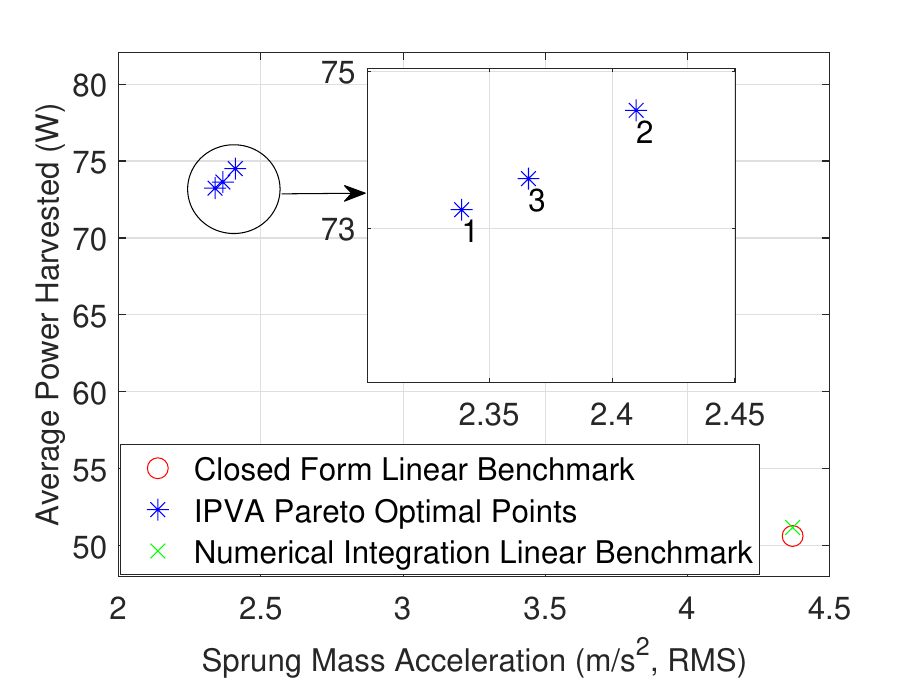}
    \caption{Pareto optimal points for the proposed IPVA-integrated system (black asterisk) and the linear benchmark on Class-C road (red circle: closed form solution, green cross: numerical integration). $\omega_0=14.83$ rad/sec.}
    \label{fig:nonvslin}
\end{figure}

\begin{figure}[!h]
    \centering
     \includegraphics[width=0.9\linewidth]{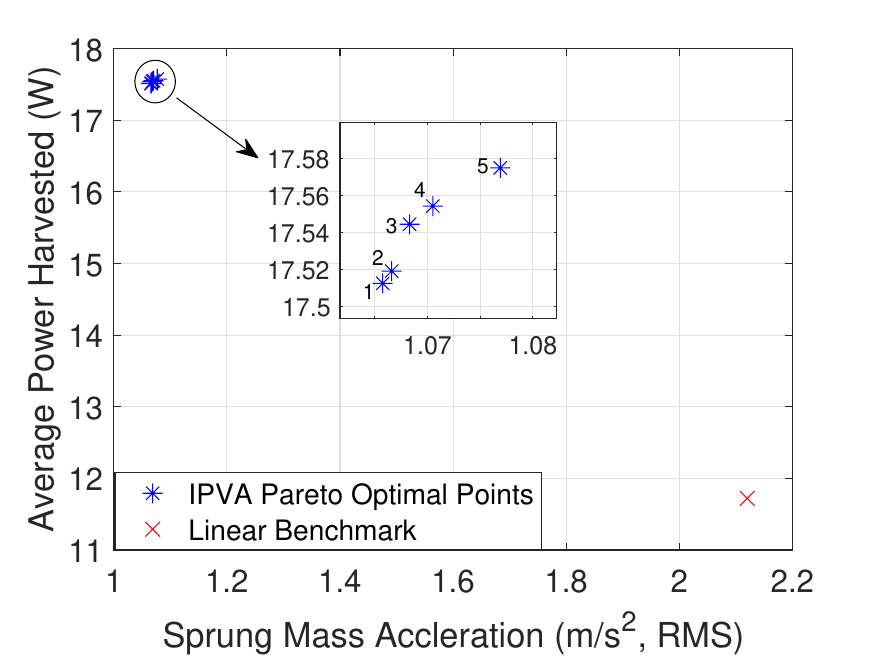}
    \caption{\textcolor{black}{Pareto optimal points for the proposed IPVA-integrated system (black asterisk) and the linear benchmark on Class-B road. }}
    \label{fig:pareto_typeb}
\end{figure}

The optimal design of the linear benchmark can be similarly obtained by solving the following optimization problem with respect to the variable $c_e$:
\begin{equation}\label{eq:optlin}
   \begin{matrix}
  \underset{{{c}_{e}}}{\mathop{\min .}}\,\text{    }\left[ \begin{matrix}
   \frac{1}{N}\sum\limits_{t=0}^{N}{-{{c}_{e}}\dot{\theta }(t)}^{2}\text{,} & \sum\limits_{t=0}^{N}{\sqrt{\frac{1}{N}(\ddot{x}_s)^2}}  \\
\end{matrix} \right]^T \\ 
  \text{s}\text{.t}\text{.   }(\ref{lb1})-(\ref{lb2}),\text{ }{{\xi }_{e}}<1.\text{  }
\end{matrix}
\end{equation}
Note that the optimization problems in (\ref{eq:optnon}) and (\ref{eq:optlin}) are computationally hard to solve directly. Alternatively, we utilize a grid search method where we discretize the optimization space into grid points and evaluate the performance of each point through  Monte-Carlo simulations. More specifically, for each grid point that corresponds to a combination of the parameters to be optimized, we run $N$ ($N=50$) simulations for sufficiently long time, each of which is based on one random generated road profile corresponding to the ISO 8608 Class-C and \textcolor{black}{Class-B} road \cite{zuo2013energy}.
Figs.~\ref{fig:nonvslin} and \textcolor{black}{\ref{fig:pareto_typeb}} summarizes the optimization results where only the Pareto optimal points are shown. \textcolor{black}{Note that instead of combining the harvested energy and ride comfort metrics in a weighted sum, the adopted Monte Carlo method allows us to inspect the Pareto optimal designs to hand-pick the one that achieves the most desirable tradeoff.}

Note in \textcolor{black}{Fig.~\ref{fig:nonvslin}}, the linear benchmark case, the Pareto front turns out to be a single point. Based on the Pareto front, we choose Point 3 as the ``optimal'' parameter set since we believe it represents the best trade-off between ride comfort and harvested energy.  This choice corresponds to the following parameters: $R_p=0.117\,m,\;r=0.0897\,m,\text{ }{{c}_{e}}=0.225\,N\cdot s\cdot m$. The optimal design for the linear benchmark corresponds to $\text{ }{{c}_{e}}=0.225\,N\cdot s\cdot m$. It can be seen in Fig.~\ref{fig:nonvslin} that the optimal IPVA system significantly outperforms the optimal linear benchmark design with about $45\%$ increase in the harvested power and $45\%$ reduction in the RMS value of sprung mass acceleration (better ride comfort). This clearly shows that our new nonlinear IPVA design can simultaneously achieve significantly better energy harvesting and ride comfort compared to conventional linear designs. \textcolor{black}{The same arguments are true for the Class-B road analysis shown in Fig.~\ref{fig:pareto_typeb} where we can choose any design point from 1 to 5 for the optimal operation of the IPVA, e.g., point 3 corresponds to design parameters $R_p=0.132\,m,\;r=0.1012\,m,\text{ }{{c}_{e}}=0.225\,N\cdot s\cdot m$ and the optimal design for the linear benchmark corresponds to $\text{ }{{c}_{e}}=0.225\,N\cdot s\cdot m$.}

\begin{figure}[!h]
    \centering
    \includegraphics[width=0.9\linewidth]{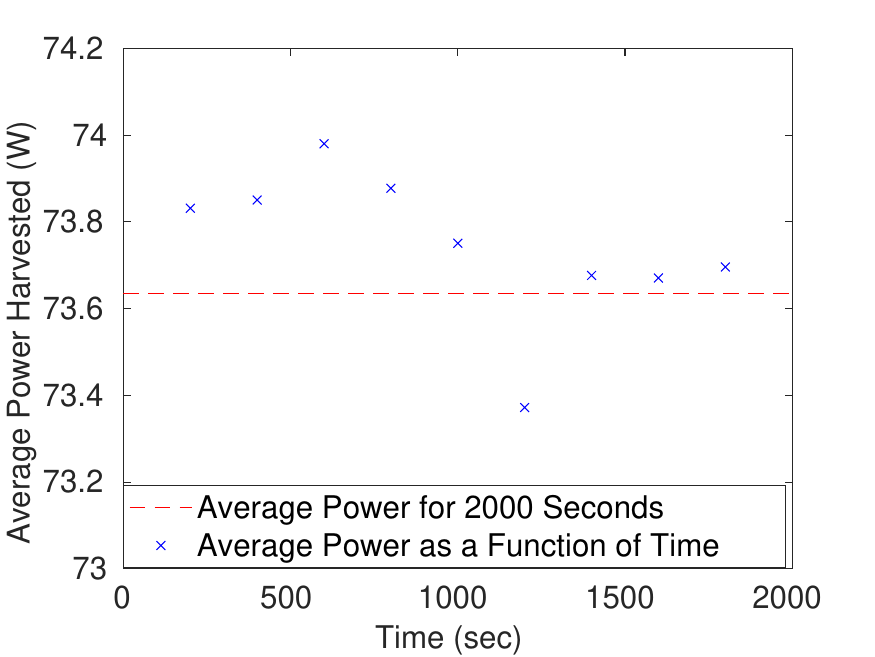}
    \caption{Average power as a function of time for Pareto point 3 in Fig.~\ref{fig:nonvslin} (black crosses) and average power over 2000 seconds (red dashed line). $\omega_0=14.83$ rad/sec.}
    \label{fig:convergence}
\end{figure}

To verify that the predictions of the average power associated with the IPVA-integrated system shown in Fig.~\ref{fig:nonvslin} are weakly stationary, the average power  as a function of time is plotted in Fig.~\ref{fig:convergence} for  the optimal parameters (Pareto point 3 in Fig.~\ref{fig:nonvslin}). It is noted that after about 1,200 seconds of integration time, the average power remains within 0.2\% of the average value for an integration time of 2,000 seconds, implying that it is approaching weak stationarity.  In addition, Fig.~\ref{fig:nonvslin} also shows the closed-form solution for RMS acceleration and average power harvested associated with the linear benchmark (see Appendix for the derivations). The solution given by numerical integration of (\ref{accelsol}) and (\ref{rms power}) for 2000 seconds, averaged over 50 realizations, is shown in Fig.~\ref{fig:nonvslin} as well. It is clear that the closed-form solution and the numerical integration solutions for the linear benchmark are very close, which further confirms that the solution given by numerical integration is very close to stationarity. 

\begin{figure}[!h]
    \centering
    \includegraphics[width=0.85\linewidth]{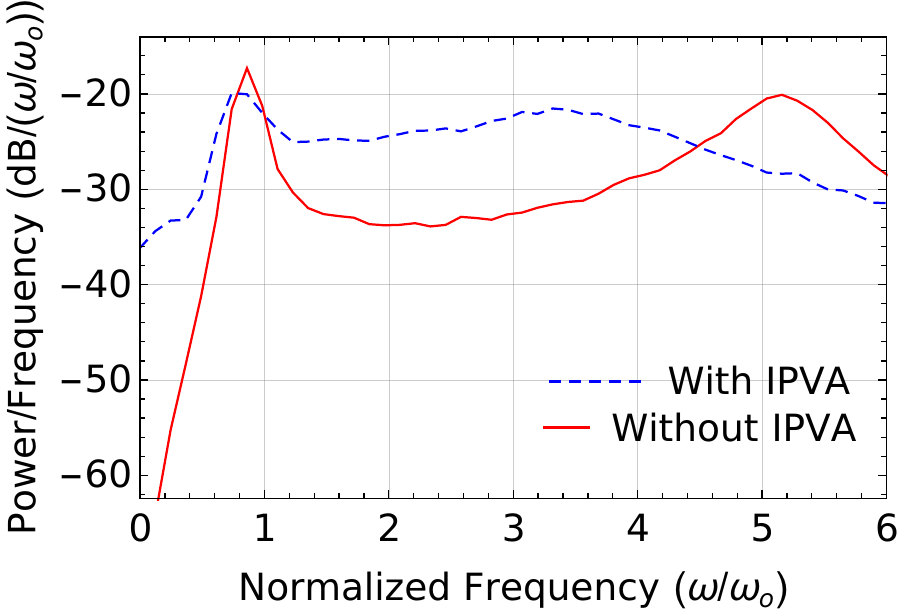}
    \caption{Power spectral density for instantaneous power with normalized frequency $\omega/\omega_0$; Linear system: solid line, IPVA-system: dashed line.}
    \label{fig:powerpsd}
\end{figure}

\begin{figure}[!h]
    \centering
    \includegraphics[width=0.85\linewidth]{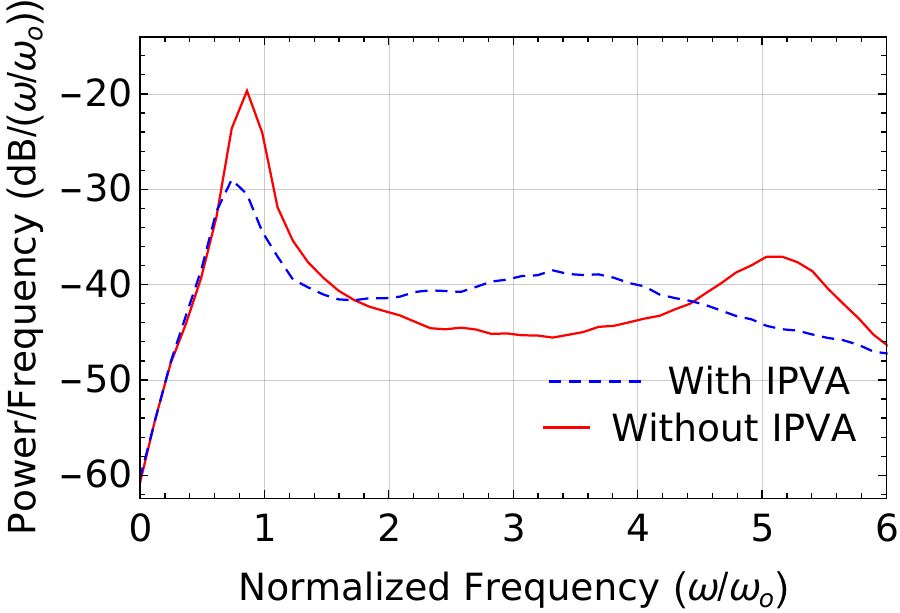}
    \caption{Power spectral density for sprung mass acceleration with normalized frequency $\omega/\omega_0$; Linear system: solid line, IPVA-system: dashed line.}
    \label{fig:accpsd}
\end{figure}

In order to further explain the performance improvement with the implementation of the IPVA, the power spectral densities (PSD's) associated with the sprung mass acceleration and instantaneous power for the IPVA-system and the linear system are numerically calculated and compared for the third Pareto point. This is shown in Figs. \ref{fig:powerpsd} and \ref{fig:accpsd}.  Note that the PSD's are a function of normalized frequency $\omega/\omega_0$, where $\omega_0=\sqrt{k_s/M_s}$. Furthermore, the natural frequencies associated with linear system were analytically calculated as $\omega_{n1}=.85\omega_0$ and $\omega_{n2}=5.18\omega_0$, which naturally correspond to the frequencies associated with the PSD peaks of linear system; see Figs. \ref{fig:powerpsd} and \ref{fig:accpsd}.
Figure \ref{fig:powerpsd} shows that the addition of the IPVA produces a super-harmonic peak at about four times the first natural frequency associated with the linear system, while the second natural frequency is nearly destroyed. Specifically, between the first natural frequency ($\omega\approx0.85\omega_0$) and super-harmonic frequency ($\omega\approx4\times0.85\omega_0$), the power spectral density with the IPVA is significantly larger, leading to greater average power relative to the linear system.  Moreover, the acceleration power spectral density for the IPVA-system shown in Fig.~\ref{fig:accpsd} displays the same super-harmonic characteristics, with negligible second natural frequency contribution, while the first peak is greatly diminished, and the super-harmonic peak is relatively small compared to second natural frequency contribution associated with the linear system. This naturally leads to lower overall acceleration for the sprung mass.

\textcolor{black}{This section is concluded with a brief discussion on the limitations of the proposed IPVA. The IPVA employs a ball-screw design and has similar limitations with other ball-screw-based EM dampers. It is known that ball-screws have a relatively lower conversion efficiency and relatively higher cost than rack-pinions \cite{abdelkareem2018}, and may have a risk of structural failure (buckling) \cite{hoo2013investigation}. Furthermore, the ball recirculating system and the planetary gear system may have reliability issues due to having complicated transmission mechanisms, which may limit the applicability to heavy-duty vehicles.}

\section{Model Predictive Controller Designs}
\textcolor{black}{In this section, we present controller designs for the IPVA-integrated suspension system to further improve its performance. Specifically, we first exploit a nonlinear MPC approach based on the derived nonlinear dynamics (subsection III-A). Furthermore, we develop a novel stochastic linearized MPC where we use a stochastically linearized model that can accurately approximate  the nonlinear dynamics while leading to much more efficient computations (subsection III-B).}
\subsection{ Nonlinear MPC (NMPC) design }
Our NMPC design is based on the derived nonlinear dynamics in (\ref{eq:EOM})-(\ref{eq:EOM3}). By defining the states
${{x}_{1}}=\theta ,\text{ }{{x}_{2}}=\dot{\theta },\text{ }{{x}_{3}}=\phi ,\text{ }{{x}_{4}}=\dot{\phi },\text{ }{{x}_{5}}={{x}_{us}},\text{ }{{x}_{6}}={{\dot{x}}_{us}}$,  the control input $u=c_e$ (electrical damping coefficient), and the external disturbance $w={{x}_{r}}$, the state-space model can be written as:
\begin{equation}
    G(x)\dot {x}=F(x,u,w),
    \label{Eq:kavaguchi}
\end{equation}    

where 
\begin{equation*} 
   G(x)= \left[ \begin{matrix}
   1 & 0 & 0 & 0 & 0 & 0  \\
   0 & {{G}_{22}} & 0 & {{G}_{24}} & 0 & M_sR  \\
   0 & 0 & 1 & 0 & 0 & 0  \\
   0 & {G_{42}} & 0 & G_{44} & 0 & 0  \\
   0 & 0 & 0 & 0 & 1 & 0  \\
   0 & M_sR & 0 & 0 & 0 & M_s+{{M}_{us}}
\end{matrix} \right],
\end{equation*}
with $G_{22}$, $G_{24}$, $G_{42}$ and $G_{44}$ being introduced in (\ref{eq:EOM})-(\ref{eq:EOM3}); and 
\begin{equation}
    F(x,u,w)= \left[\begin{matrix}
  {{x}_{2}} \\ 
 F_2(x) \\
  {{x}_{4}} \\ 
  -u(x_4-x_2)-mR_{p}r\sin(x_3)x_2^2-k_{p}x_3 \\ 
 {{x}_{6}} \\ 
  -{{k}_{t}}({{x}_{5}}-w)
\end{matrix} \right], \notag
\end{equation}
where
\begin{equation}
\centering
\begin{aligned}
   F_2(x)&=-{{c}_{m}}{{R}^{2}}{{x}_{2}}+{u}({{x}_{4}}-{{x}_{2}})-k{{R}^{2}}{{x}_{1}}\\&+2m{{R}_{p}}r{{x}_{4}}{{x}_{2}}\sin ({{x}_{3}})+m{{R}_{p}}r\sin ({{x}_{3}}){{x}_{4}}^{2}.
\end{aligned}
\end{equation}

 It can be shown that $G(x)$ is positive definite and therefore invertible. The objectives of the control designs are twofold: 1) Achieve good ride comfort by minimizing the sprung mass acceleration (i.e., $\ddot{x}_{s}=\ddot{{x}}_{us}+R\ddot{\theta}=\dot{x}_6+R\dot{x}_2$); and 2) Realize efficient energy harvesting by maximizing the regenerated power (i.e., $P=c_e(\dot{\theta}-\dot{\phi})^2=u(x_2-x_4)^2$) by controlling the damping $u$ in real time \cite{chen2020mpcbased}. Hence, the continuous time economic stage cost function can be defined as
 \begin{equation}
     l\left( x,u \right)={{\alpha }_{1}}{{\left( {{{\dot{x}}}_{6}}+R{{{\dot{x}}}_{2}} \right)}^{2}}-{{\alpha }_{2}}u{{\left( {{x}_{2}}-{{x}_{4}} \right)}^{2}},
     \label{eq:stage_cost}
 \end{equation}
 where $\alpha_1$ and $\alpha_2$ are positive weighting factors to tradeoff the two objectives. As such, the NMPC problem can be formulated as:
 \begin{equation}\label{NMPC}
\begin{array}{l}
\min_{U} \quad J=\sum_{k=0}^{N-1} l_{d}(x(k),u(k)) \\
\\
\text { s.t. } \quad x(0)=x_{0} ,\qquad u(k) \in \mathbb{U}_{d}, \\
\\
x(k+1)=\left[G(x(k))^{-1} F(x(k), u(k), w(k))\right]_{d},
\end{array}
\end{equation}
where $U=[u(0),\cdots,u(N-1)]$ is the optimization variable, $l_d$ is the discrete stage cost of $l$, $N$ is the prediction horizon, and $\mathbb{U}_{d}$ represents the control constraints.

Note that this NMPC problem is computationally heavy and is difficult for onboard implementation due to fast dynamics of suspension systems. Therefore, we next present a sub-optimal MPC that is computationally efficient and thus more suitable for practical uses.

\subsection{ Stochastic linearized MPC (SL-MPC) }
\begin{figure}[!h]
    \centering
    \includegraphics[width=.9\linewidth]{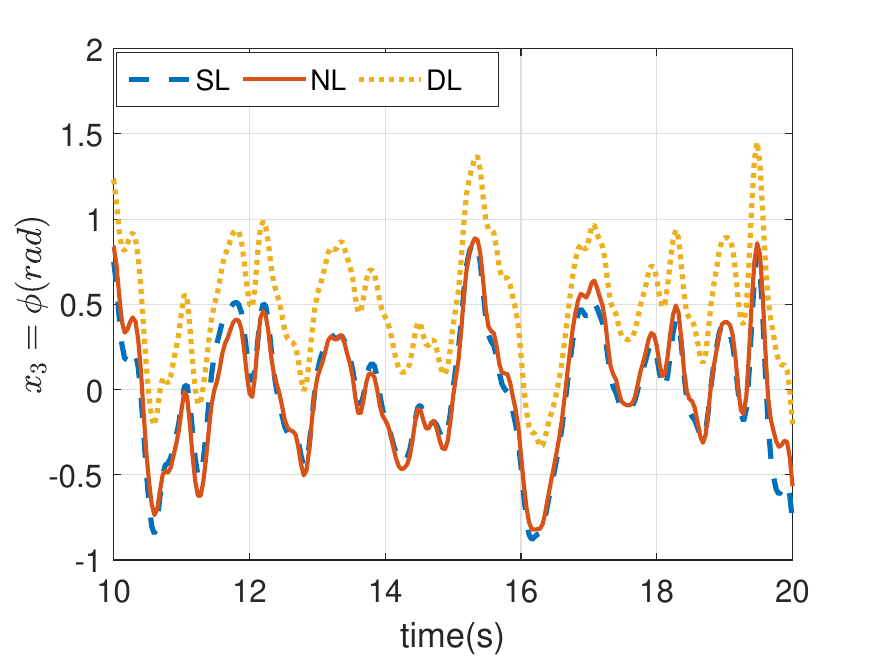}
    \caption{Accuracy of stochastic linearization (SL) approach relative to deterministic linearization (DL).}
    \label{nlsl}
\end{figure}

As the obtained dynamics of the IPVA-integrated suspension system is inherently nonlinear, the NMPC formulation above is computationally expensive and difficult for online implementations. It would be meritorious if we can find a good linear approximation of the nonlinear dynamics to achieve efficient computations. One option is to linearize the nonlinear system around equilibrium points, referred to as conventional (or deterministic) linearization and it is shown that this approach does not work well for the considered nonlinear system as it produces large prediction error compared to the original nonlinear system (see e.g., Figure~\ref{nlsl}). In this subsection, we propose a stochastic linearization approach where we derive a linear approximation such that its system response is statistically close to the nonlinear response when subject to external random excitation. This is especially appealing as our EHSA system is indeed subject to random road disturbance.  The stochastic linearization approach is a powerful and efficient tool for capturing the complex and random behavior of a nonlinear system \cite{roberts2003random}. The goal is to obtain a linear system such that its deviation from the original nonlinear system is small in the sense of expectations. 
To this end, we first write the EOM of the IPVA dynamics (\ref{eq:EOM})-(\ref{eq:EOM3}) in an alternative form as: 
\begin{equation}\label{eq:EOMgen.}
  \mathbf{M}_l\mathbf{\ddot{q}}+\mathbf{C}_l\mathbf{\dot{q}}+\mathbf{K}_l\mathbf{q}+\mathbf{\Phi }(\mathbf{q,\dot{q},\ddot{q}})=\mathbf{Q}(t),
 \end{equation}
 where we have
 \[\mathbf{M}_l = 
 \begin{bmatrix}
    M_s R^2+ J+ mR_p^2 + mr^2 & mr^2+J_p & RM_s\\
    mr^2+J_p & mr^2+J_p & 0 \\
    RM_s & 0 & M_s+M_{us}
 \end{bmatrix},\]
\[ \boldsymbol{\Phi} = \begin{bmatrix}
    \mathbf{\Phi_1} & \mathbf{\Phi_2} & \mathbf{\Phi_3}
 \end{bmatrix}^T,\]
\[ \mathbf{C}_l = \begin{bmatrix}
    c_m R^2+c_e & -c_e  & 0\\
    -c_e & c_e & 0 \\
    0 & 0 & 0
 \end{bmatrix}, \ 
 \mathbf{K}_l = \begin{bmatrix}
    k_s R^2 & 0 & 0 \\
    0 & 0 & 0 \\
    0 & 0 & k_t
 \end{bmatrix},\\
 \]

\begin{equation}
 \begin{matrix}
  \label{eq:paras}
  {{\mathbf{\Phi }}_{\mathbf{1}}}=2m{{R}_{p}}r\cos (\phi )\ddot{\theta }+m{{R}_{p}}r\cos (\phi )\ddot{\phi }-2m{{R}_{p}}r\dot{\phi }\dot{\theta }\sin (\phi )\\-m{{R}_{p}}r\sin (\phi ){{{\dot{\phi }}}^{2}}, \\
  {{\mathbf{\Phi }}_{\mathbf{2}}}=m{{R}_{p}}r\cos (\phi )\ddot{\theta }+m{{R}_{p}}r\sin (\phi ){{{\dot{\theta }}}^{2}}, \
  {{\mathbf{\Phi }}_{3}}=0.
\end{matrix}
\end{equation}

 The matrices $\mathbf{M}_l$, $\mathbf{C}_l$ and $\mathbf{K}_l$ are the linear inertia, damping, and stiffness matrices, respectively;  $\mathbf{\Phi }$ is the collection of nonlinear terms in the equations; and $\mathbf{Q}(t)$ is the generalized force. The main idea in SL approach is to find equivalent deterministic inertia, damping, and stiffness matrices (denoted by $\mathbf{M}_e$, $\mathbf{C}_e$ and $\mathbf{K}_e$, respectively) such that when replaced by the nonlinear terms, the system responses to the random disturbance signal are similar in the statistical sense, that is,  the following equivalent stochastic linearized system
\begin{equation}\label{eq:SLEOM}
    (\mathbf{M}_l+{{\mathbf{M}}_{e}})\mathbf{\ddot{q}}+(\mathbf{C}_l+{{\mathbf{C}}_{e}})\mathbf{\dot{q}}+(\mathbf{K}_l+{{\mathbf{K}}_{e}})\mathbf{q}=\mathbf{Q}(t)
\end{equation}
is close to (\ref{eq:EOMgen.}). More specifically,  we seek  $\mathbf{M_e}$, $\mathbf{C_e}$ and $\mathbf{K_e}$ by solving the following optimization problem:
\begin{equation}
\label{eq:epsopt}
  \underset{{{\mathbf{M}}_{e}},{{\mathbf{C}}_{e}},{{\mathbf{K}}_{e}}}{\mathop{\min .}}\,E\{{{\mathbf{\varepsilon }}^{T}}\mathbf{\varepsilon }\}
\end{equation}
where 
\begin{equation}
    \mathbf{\varepsilon}=\mathbf{\Phi }(\mathbf{q,\dot{q},\ddot{q}})-{{\mathbf{M}}_{e}}\mathbf{\ddot{q}}-{{\mathbf{C}}_{e}}\mathbf{\dot{q}}-{{\mathbf{K}}_{e}}\mathbf{q}
\end{equation}
is the $n-$dimensional vector difference between the actual nonlinear system and the stochastically linearized system  \cite{roberts2003random}, i.e., the difference between (\ref{eq:EOMgen.}) and (\ref{eq:SLEOM}).

Assuming the road disturbance is Gaussian and by following the stochastic optimization procedures in \cite{roberts2003random}, one can find the elements of $\mathbf{M_e}$,
$\mathbf{C_e}$ and $\mathbf{K_e}$ as:
\begin{align}\label{eq:me}
  & {{\mathbf{M}}_{e,ij}}=E\{\frac{\partial {{\mathbf{\Phi }}_{i}}}{\partial {{{\mathbf{\ddot{q}}}}_{j}}}\}, \\ 
 & {{\mathbf{C}}_{e,ij}}=E\{\frac{\partial {{\mathbf{\Phi }}_{i}}}{\partial {{{\mathbf{\dot{q}}}}_{j}}}\}, \\ 
 & {{\mathbf{K}}_{e,ij}}=E\{\frac{\partial {{\mathbf{\Phi }}_{i}}}{\partial {{\mathbf{q}}_{j}}}\},\text{    }\quad {\text for }\;i,j=1,2,3, \label{eq:ke}
\end{align}
where $\mathbf{q}_{1}=\theta$, $\mathbf{q}_{2}=\phi$,  $\mathbf{q}_{3}=x_{us}$, and $\Phi_1$, $\Phi_2$, and $\Phi_3$ are introduced in Eqn. \ref{eq:paras}. Solving Eqns. (\ref{eq:me})-(\ref{eq:ke}) can be done analytically or by Monte-Carlo simulation to calculate the expectations. The former poses a challenge due to the complication of nonlinear terms whereas the latter could be done with desired accuracy with adequate number of simulations for any type of nonlinear terms and road disturbance signal. In this paper, we exploit the latter approach to obtain the $\mathbf{M_e}$, $\mathbf{C_e}$, and $\mathbf{K_e}$ terms.

The system responses of the stochastic linearized (SL) system and the conventional deterministic linearized (DL) system is compared with the original nonlinear (NL) system, where the results for $x_3$ on one sample road profile is shown in Fig.~\ref{nlsl}. It is clear that the stochastic linearization is a much closer representation of the nonlinear dynamics.

While the SL approach discussed above can generate a linear model that closely approximates the nonlinear dynamics, the optimization in Eqn.~\ref{eq:epsopt} has no guarantees in controllability or stabilizability, which is a prerequisite for control designs. In case that the obtained SL system is not stabilizable, a constrained optimization problem can be formulated to guarantee stabilizability. The process is detailed in Appendix-B where the essential idea is to find stabilizable matrices that are close to the original SL matrices, e.g., in the sense of matrix norm. 

With the stochastically linearized system, one can formulate the following SL-MPC problem:
\begin{equation}\label{SLMPC}
\begin{aligned}
&\qquad\qquad\min_{\mathbf{U}} \sum_{k=0}^{N-1} l_{d}(x(k),u(k)) \\
&\text {subject to } \quad x(0)=x_{0} ,\qquad u(k) \in \mathbb{U}_{d}, \\
&x(k+1)=A_{s}x(k)+B_{s}x(k)u(k)+D_{s}w(k),
\end{aligned}
\end{equation}
where $\mathbf{U}=\left[u(0);\cdots;u(k-1)\right]$, and $A_{s}, B_{s}$ and $D_{s}$ can be found by inspecting (\ref{eq:SLEOM}) with $x=[\mathbf{q},\mathbf{\dot{q}}]^T$. Note that the system model in (\ref{SLMPC}) has a bilinear term, $B_{s}x(k)u(k)$, where the state and the control variable are cross multiplied. We next follow the reformulation approach in \cite{savaresi2010semi} to transform this bilinear model to a linear one. In specific, we substitute the damping force term, $u(x_4-x_2)$ by $F_d$, which follows by changing the control variable from $u$ to $F_d$ and introducing passivity constraints to the model, i.e.,
\begin{equation}
\label{eq:LTI}
  \begin{aligned}
  &x(k+1)={{A}_{l}}x(k)+{{B}_{l}}{{F}_{d}}(k)+{{D}_{l}}w(k) \\ 
  &\text{s}\text{.t}\text{.  }-{{F}_{d}}(k)({{x}_{4}}(k)-{{x}_{2}}(k))\le 0, \\
  &\left[F_{d}(k)-{{c}_{\max }}({{x}_{4}}(k)-{{x}_{2}}(k))\right]\cdot F_{d}(k)\le 0,\text{      } \\ 
\end{aligned} 
\end{equation}
where $A_l$, $B_l$, and $D_l$ are the induced matrices after the new control formulation. This linear time invariant form will make the MPC problem much more computationally efficient for computations. 

\begin{figure}[t]
    \centering
    \includegraphics[width=0.95\linewidth]{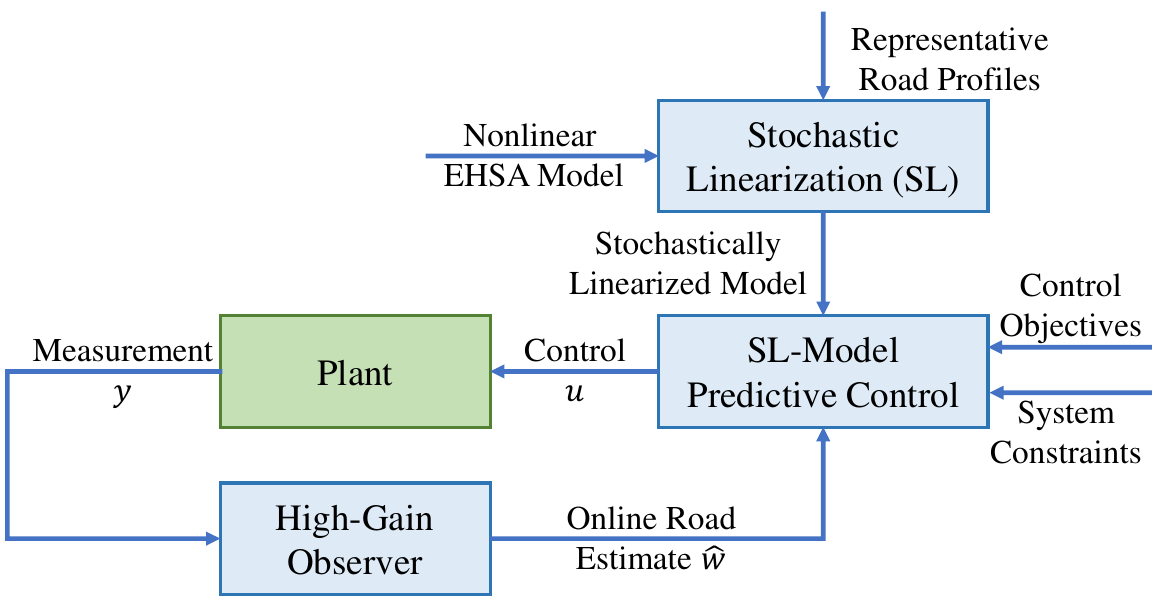}
    \caption{\textcolor{black}{Flowchart of SL-MPC.}}
    \label{fig:flowchart}
\end{figure}

\textcolor{black}{We would like to wrap up this section with a flow chart to summarize the work flow, modeling assembly and the control algorithm in SL-MPC. As shown in Figure \ref{fig:flowchart}, our proposed framework starts with the \textit{Stochastic Linearization} (SL) of the Nonlinear EHSA model given the \textit{Representative Road Profile}. This stochastic linearized model is then used for the MPC task by considering the \textit{Control Objectives} (i.e., the trade off between ride comfort, energy harvesting, and vehicle handling) and \textit{System Constraints} (i.e., electrical damping maximum value). These steps form the SL-Model Predictive Control block in the flowchart. The obtained control command is then used for controlling the \textit{Plant}. Lastly, the measurement from plant is used by the \textit{High-Gain Observer} to have an estimate of the road profile which is used by the SL-MPC as the road preview information.}

\section{Simulation Results}
In this section, extensive simulations are presented to investigate the performance of different control designs including  the passive IPVA system, NMPC, and SLMPC. Furthermore, we evaluate the control designs in different road preview settings.
\textcolor{black}{The road preview refers to the availability of road profile signal data, from time step $t$ to the end of the horizon, $t+N-1$, i.e., $\left(w(t), w(t+1),...,w(t+N-1)\right)$, for the MPC task. Specifically, the MPC at each time step $t$ solves a constrained optimization problem with a pre-defined objective and horizon $N$ by predicting the trajectory of the system from time step $t$ to $t+N-1$ using the system dynamics (e.g., Eqn.~\ref{NMPC}) and the road profile information $\left(w(t), w(t+1),...,w(t+N-1)\right)$, if available. It is thus clear that a better estimate of the road profile will lead to a more accurate prediction, and subsequently better control performance.}
In the first setting, we assume there is a complete road preview  e.g., obtained from recent road information estimation techniques \cite{li33,li34}. In the second setting,  we exploit an online road profile estimation algorithm, and use the last road disturbance estimation (LRDE) while keeping it constant across the prediction horizon. This is done using the HGO design introduced in Appendix-C. In the third setting, we use a noisy version of the preview with different signal to noise (SNR) ratios in the prediction horizon to evaluate the performance. Three different SNR values are considered which corresponds to road profiles with large noise (SNR 10), moderate noise (SNR 15), and small noise (SNR 20). The first setting can be used to evaluate the performance cap while the second and third are realistic settings that has practical implications.  The road disturbance signals follow a Class-C (average) road, which are generated following the procedure outlined in \cite{zuo2013energy} and the vehicle speed is considered as 60 mph.
The system parameters for simulation are summarized in Table~\ref{paramet}.
\begin{table}[!h]
\begin{center}
\begin{tabular}{|c||c||c||c|}
\hline 
$M_s$ & $M_{u }$ & $k_{t}, k_{s}$ & $k_{p}$ \\
\hline
$250\text{ }kg$ & $35\text{ }kg$ & $150,55 \text{ }kN$ & $0.852\text{ }N$ \\
\hline
$J_r$ & $J$ & $J_h$ & $J_p$ \\
\hline
$1.21\times 10^{-4} kg\cdot m^2$ & 0 & 0 & 0 \\
\hline
$m$ & $N$ & $[u_{min},u_{max}]$ & $T_{s}$ \\
\hline
$2.5\text{ }kg$ & $15$ & $[{0,0.225}]$ & $0.01\text{ }s$ \\
\hline
\end{tabular}
\end{center}
\caption{Simulation parameters}
\label{paramet}
\end{table}

\begin{figure}[!h]
    \centering
    \includegraphics[width=1\linewidth]{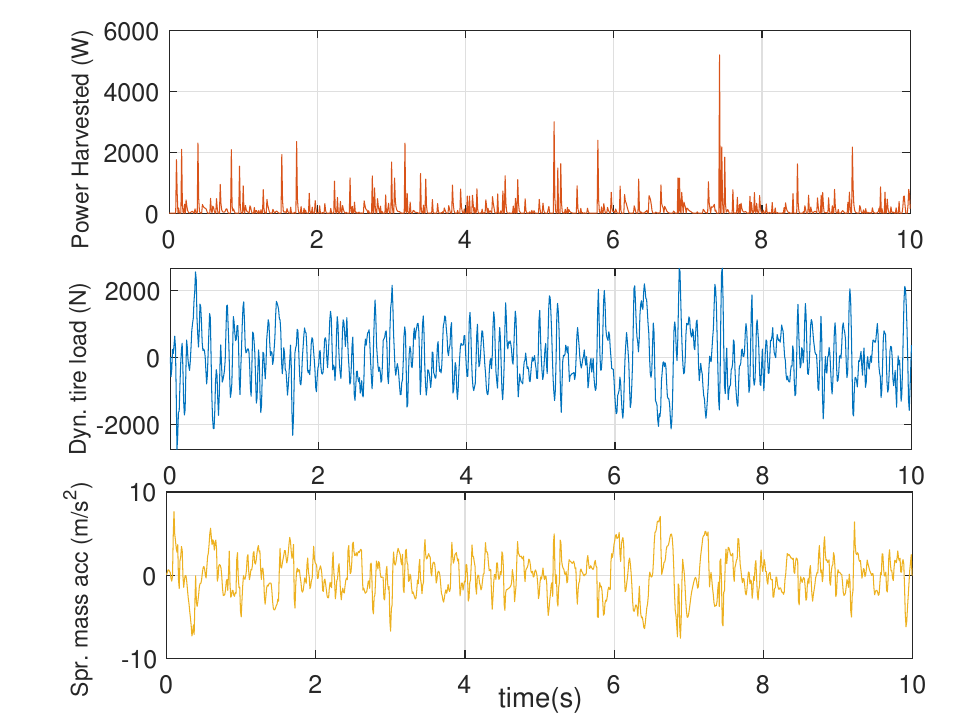}
    \caption{Power harvested, dynamic tire load and sprung mass acceleration as a function of time for one sample of IPVA operation with SL-MPC control}
    \label{fig:time_domain_plts}
\end{figure}

\textcolor{black}{A sample episode of harvested power,  dynamic tire load and sprung mass acceleration as functions of time are shown in Figure \ref{fig:time_domain_plts}.}

\begin{figure}[!h]
    \centering
    \includegraphics[width=1\linewidth]{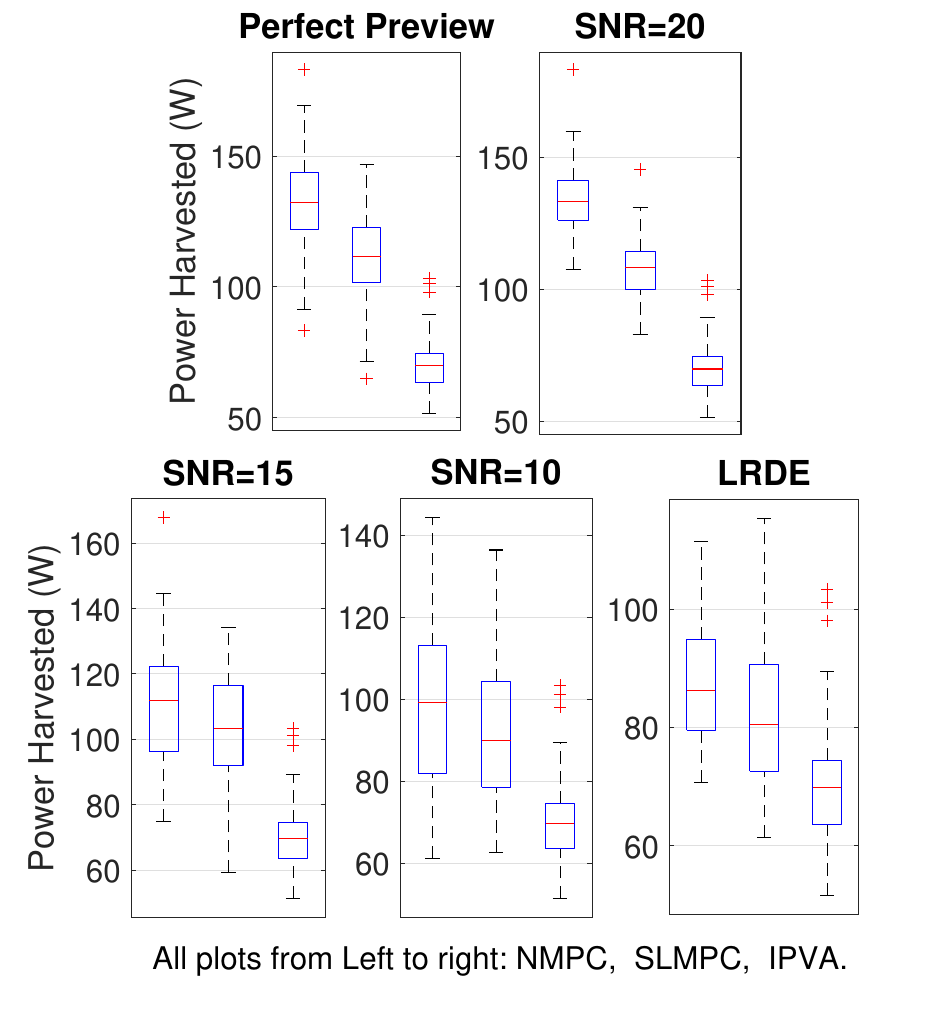}
    \caption{Box plot of control performance comparison for the case of maximizing the energy  harvesting. Red horizontal lines represent the mean and the box heights represent the standard deviations. }
    \label{maxenergyDIS}
\end{figure}

\textcolor{black}{\subsection{Maximizing harvested energy}}
We first examine the case of maximizing the power harvested where we choose $\alpha_1=0$ and $\alpha_2=1$.  The results are summarized in Fig.~\ref{maxenergyDIS}, which shows that both NMPC and SLMPC significantly increase the average power harvested  across all cases. \textcolor{black}{More specifically, for the perfect preview case, the average power harvested for NMPC and SL-MPC is increased nearly 91$\%$ and 60$\%$, respectively. These numbers for the cases where we have an SNR = 20,15,10 or we use the LRDE as the preview for the entire horizon are, respectively, 90$\%$ and 56$\%$, 60 $\%$ and 49$\%$, 43$\%$ and 28$\%$, 24$\%$ and 15$\%$}.  It can be seen that using stochastically linearized model does not cause a major performance degradation but leads to much greater computational efficiency as will be shown later.
\begin{figure}[!t]
    \centering
    \includegraphics[width=0.9\linewidth]{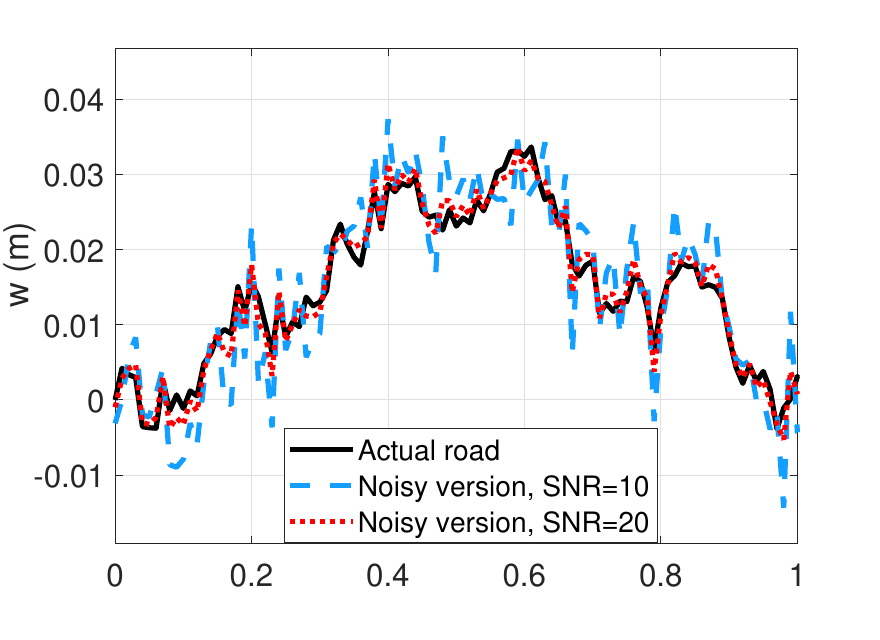}
    \caption{Actual road profile and the noisy version of road profile with different SNR levels. }
    \label{RPs}
\end{figure}

\color{black}
{\subsubsection{Efficiency Analysis}}
An efficiency analysis is used to further quantify the energy harvester performance. To accomplish this, we start with the derivation of mechanical and electrical efficiency associated with the linear system, followed by the IPVA system.

Denote mechanical efficiency as $\eta_{m,l}$, mean input power as $\left\langle P_{in}\right\rangle$, and mean output power as $\left\langle P_{out}\right\rangle$ where $\left\langle \cdot \right\rangle$ denotes statistical expectation. Note that $\left\langle P_{out}\right\rangle=P$ is defined in the appendix. The total input power can be calculated as the sum of the mechanical power input to the ball screw $\left\langle P_{in}\right\rangle$ and power lost due to mechanical damping $c_m$, $\left\langle P_{Lost} \right\rangle$. Note that the latter is introduced to account for mechanical energy loss due to the ball-screw and generator gears. The total input power to the ball screw is simply the product of the output torque ($T$) and angular velocity of the generator ($T\dot{\theta}$). The equation of motion for the generator is given as 
\begin{equation}\label{linear generator eom}
    J_r\ddot{\theta}+c_e\dot{\theta}=T
\end{equation}
Multiplying the left hand side of Eq.~(\ref{linear generator eom}) by $\dot{\theta}$ and taking the statistical expectation will give the mean power input. Furthermore, the mean dissipated power in the damper is given by 
\begin{equation}
    \left\langle P_{Lost} \right\rangle=c_m R^2\left\langle\dot{\theta}^2\right\rangle.
\end{equation}
Finally, the sum of the input power and power dissipated is given by
\begin{equation}\label{linear input power}
    \left\langle P_{Total}\right\rangle=(c_m R^2+c_e)\left\langle\dot{\theta}^2\right\rangle+J_r\left\langle \ddot{\theta}\dot{\theta}\right\rangle.
\end{equation}
In order to calculate the second term on the right hand side of Eq.~(\ref{linear input power}), $\ddot{\theta}$ is first written in terms of state variables according to Eq.~(\ref{lb1}). We then make use of statistical moment equations to complete the computation. The reader is referred to \cite{bover_moment_1978} for a detailed explanation of these equations. In this particular case, $\left\langle \ddot{\theta}\dot{\theta}\right\rangle=0$, implying that there is no statistical correlation between the acceleration and velocity of the ball screw. Additionally, we note that $\frac{1}{c_e}\left\langle P_{out}\right\rangle=\left\langle\dot{\theta}^2\right\rangle$ and so 
\begin{equation}\label{mechanical efficiency}
    \eta_{m,l}=\frac{\left\langle P_{out}\right\rangle}{\left\langle P_{Total}\right\rangle}=\frac{c_e}{c_e+c_mR^2}.
\end{equation}
As seen in Eq.~(\ref{mechanical efficiency}), the mechanical efficiency for the linear system is dependent on electrical damping, mechanical damping and the lead of the ball screw through $R=L/2\pi$. It is worth noting that mechanical damping $c_m=148.32$ Ns/m was chosen such that the linear benchmark has a mechanical efficiency ($\eta_{m,l}\approx 60\%$) similar to the linear EM damper reported in \cite{li2012electromagnetic}. The same value of $c_m$ is employed in the IPVA system for a fair comparison.

To compute the electrical efficiency it is common to assume $c_e(t)=M\kappa_e\kappa_t/R(t)$, where $R(t)=R_{int}+R_{Load}(t)$ and $R_{int}$ is the internal resistance of the generator\cite{li2012electromagnetic}. For the purpose of this paper, $R_{int}=6.6~\Omega$, $M\kappa_e\kappa_t=11~N\cdot s\cdot m\cdot\Omega$ as directly obtained and reverse engineered from the data given in \cite{li2012electromagnetic}. The electrical efficiency is then defined as the ratio of output with zero internal resistance to total output power $\left\langle P_{out}\right\rangle$. As such, we can find
\begin{equation}\label{electrical efficiency}
    \eta_{e}=\frac{R_{Load}(t)}{R_{int}+R_{Load}(t)}=\frac{M\kappa_e\kappa_t-R_{int}c_e(t)}{M\kappa_e\kappa_t}.
\end{equation}
Note that Eq.~(\ref{electrical efficiency}) applies to the system with IPVA as well. Another remark is that electrical efficiency seems to increase with an decrease in electrical damping or internal resistance. However, an electrical damping coefficient equal to zero corresponds to the case of zero power harvested. Therefore, efficiency is set to zero for cases when $c_e(t)=0$. 

\begin{figure}[!t]
    \centering
    \includegraphics[width=0.8\linewidth]{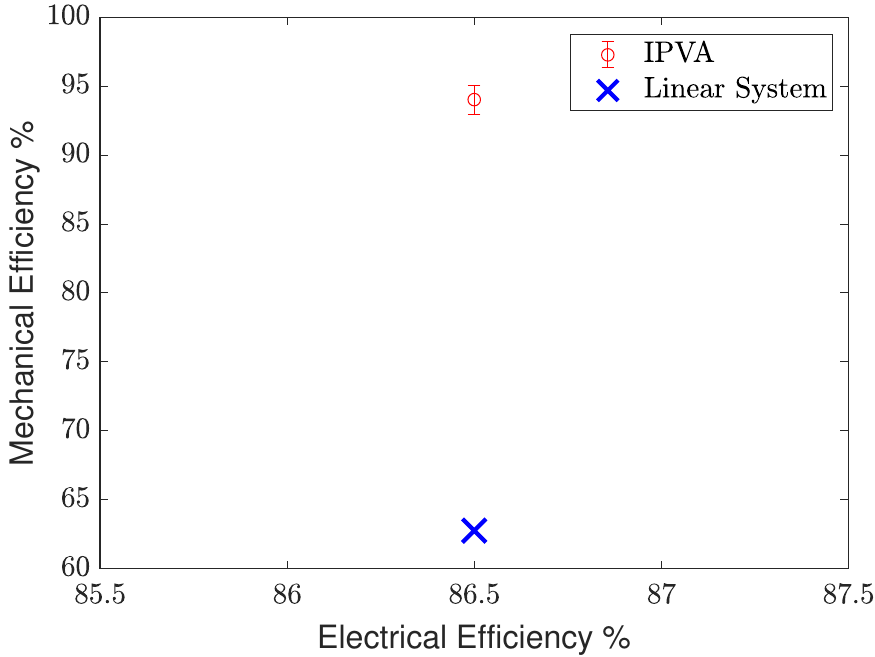}
    \caption{Mechanical and electrical efficiency for passive IPVA and passive linear system with and error bar indicating standard deviation.}
    \label{passive mechanical efficiency}
\end{figure}

In order to derive the mechanical efficiency of the IPVA system, we choose to take a Lagrangian approach with a holonomic constraint $g(\theta,x_{us},x_s)=R\theta+x_{us}-x_{s}=0.$ To this end, we first rewrite  Eq.~(\ref{Eq:lagrange}) as 
\begin{equation}
\frac{d}{d t} (\frac{\partial L}{\partial \dot{\mathbf{q}}_{i}})-\frac{\partial L}{\partial \mathbf{q}_{i}}=Q_{i}+\lambda\frac{\partial g}{\partial \mathbf{q}_{i}},
\label{Eq:lagrange with constraint}
\end{equation}
where $q_1=\theta, q_2=\phi, q_3=x_s, q_4=x_{us}$ and $\lambda$ is known as a Lagrangian multiplier while it is also the constraint force acting on the IPVA required to impose the constraint $g$. We are further required to rewrite the Lagrangian in terms of unconstrained potential and kinetic energies. To this end, we find 
\begin{equation}\begin{aligned}
T &= {T}_{M_{us}}+T_{M_s}+T_c + T_p + T_r \\ 
&={1\over 2} M_{us}\left( \dot{x}_{us}\right)^2+{1\over 2} M_s\left(\dot{x}_{s}\right)^2\\&+ {1\over 2}m\left(R_p^2\dot{\theta}^2 + r^2\left(\dot{\theta}+\dot{\phi}\right)^2+2R_pr\cos\left(\phi\right)\dot{\theta}\left(\dot{\theta} + \dot{\phi}\right)\right) \\
&+  {1\over 2} J_r  (\dot{\phi}-\dot{\theta})^2,
\label{eq:kinetic unconstrained}\end{aligned}
\end{equation}
and
\begin{equation}\label{potential unconstrained}
V = {1\over 2}k_s(x_s-x_{us})^2+{1\over 2}k_{t}(x_{us}-x_r)^2.
\end{equation}
Additionally, we rewrite virtual work as
\begin{equation}\label{virtual work constrained}
\begin{aligned}
\centering
  	\delta W_{nc} &= \delta W_r + \delta W_m =-c_e (\dot{\phi}-\dot{\theta})\cdot\delta\phi\\&+ c_e (\dot{\phi}-\dot{\theta})\cdot\delta\theta-c_m\left(\dot{x}_s-\dot{x}_{us}\right)\delta \theta.
\end{aligned}
\end{equation}
and $Q_i$ corresponds to the coefficient of $\delta q_i$ in Eq.~(\ref{virtual work constrained}).  Substituting Eq.~(\ref{virtual work constrained}), Eq.~(\ref{eq:kinetic unconstrained}) and Eq.~(\ref{potential unconstrained}) into Eq.~(\ref{Eq:lagrange with constraint}) finally allows one to relate constraint force $\lambda$ to state variables, accelerations and system parameters. Choosing the simplest relationship, corresponding to generalized coordinate $q_3=x_s$, we have
\begin{equation}\label{lambda}
    \lambda=-\left(M_s\ddot{x}_s+c_mR\dot{\theta}+kR\theta\right).
\end{equation}
Note that Eq.~(\ref{lambda}) is just a statement of Newton's second law for the sprung mass which could be expected as $\lambda$ must be equal and opposite to the constraint force imposed on the sprung mass in order to satisfy the holonomic constraint.
Next we sum the power dissipated by the mechanical damper and instantaneous input power to IPVA, where the instantaneous input power is $\lambda R\dot{\theta}$. The sum amounts to
\begin{equation}\label{IPVA input power}
    P_{Total,IPVA}=c_m R^2\dot{\theta}^2+\lambda R\dot{\theta}=-RM_s\dot{\theta}\ddot{x}_s-kR^2\theta\dot{\theta}.
\end{equation}
One remark is that the the sprung mass acceleration $\ddot{x}_s$ can be written in terms of system parameters and state variables and so this dependence is implied rather than explicitly shown in Eq.~(\ref{IPVA input power}).
The instantaneous output power going to the harvester is then simply

\begin{equation}\label{}
    P_H=c_e\left(\dot{\phi}-\dot{\theta}\right)^2.
\end{equation}
To define the mechanical efficiency in this case we choose to compute the average input and output power in each of $N=100$ realizations in the time domain, followed by the computation of efficiency in each realization and an ensemble average with standard deviation computation. The efficiency in each realization is found to be
\begin{equation}
    \eta_{m}=\frac{\int_{t=0}^{t=t_f}P_H dt}{ \int_{t=0}^{t=t_f}P_{Total,IPVA} dt},
\end{equation}
where $t_f=5$ as the efficiency value was found to be relatively fixed after 5 seconds. 
The reason for the averaging is because the calculation of power input to the IPVA may lead to negative values resulting in negative efficiencies.
The mechanical efficiency and electrical efficiency for the passive IPVA as well as linear system is shown in Fig.~\ref{passive mechanical efficiency}. 

A remark is given to the interpretation of the mechanical efficiencies shown in Fig.~\ref{passive mechanical efficiency}. It should be noted that mechanisms that are associated with the pendulums, including the planetary gear system and pendulum bearings, should lead to mechanical energy loss in practice. Therefore, the mechanical efficiency of IPVA should be lower in practice. In this regard, Fig.~\ref{passive mechanical efficiency} only shows the maximum mechanical efficiency IPVA can theoretically have provided that the IPVA and linear benchmark have the same mechanical damping coefficient $c_m=148.32$ Ns/m.

\begin{figure}[!h]
    \centering
    \includegraphics[width=1\linewidth]{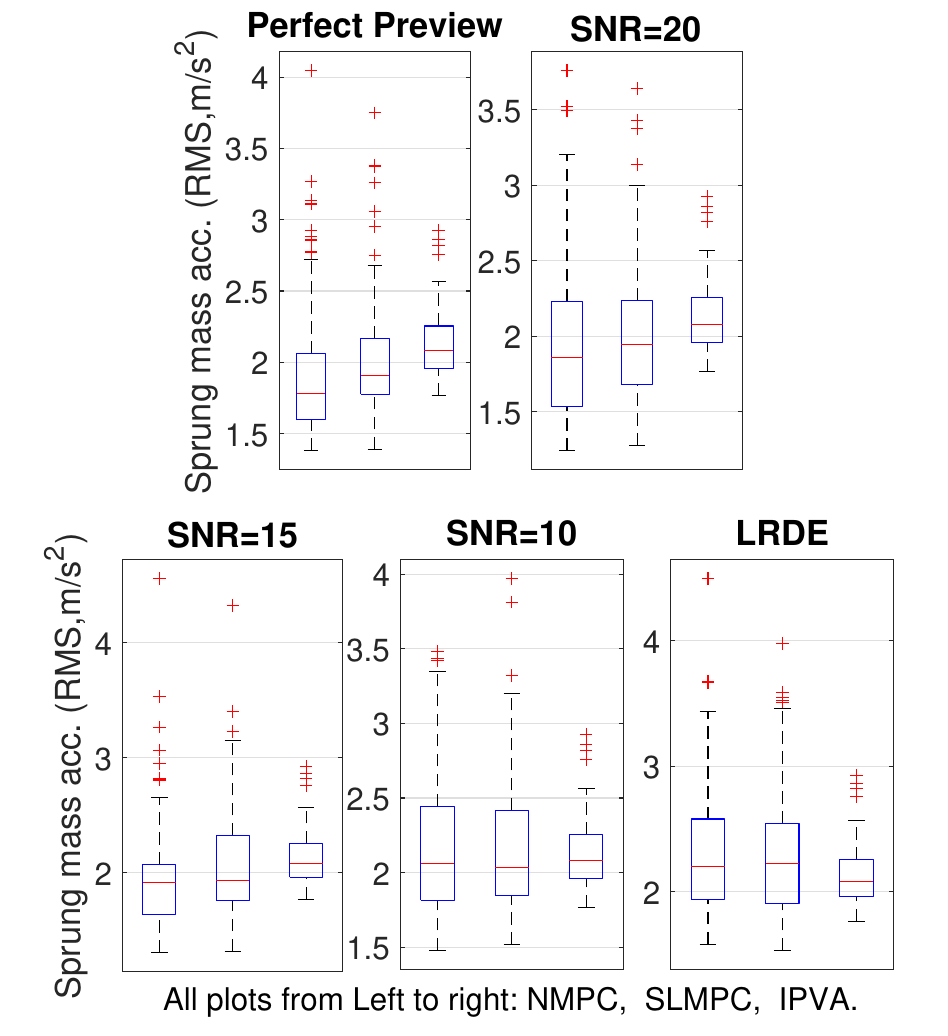}
    \caption{Box plot of control performance comparison for the case of minimizing vertical acceleration. Red horizontal lines represent the means while the box heights represent the standard deviations. }
    \label{minacc}
\end{figure}

\textcolor{black}{This section is concluded by a comparison of the simulated power of IPVA with comparable results reported in the literature; see Table~\ref{tab:Simulation_Survey}. It should be noted that only simulation results that consider the ISO 8608 road classes are included in the table for a fair comparison. The reader is referred to \cite{abdelkareem2018} for the harvested power of other EHSAs that were obtained experimentally or numerically under other excitation conditions. As shown in Table~\ref{tab:Simulation_Survey}, the IPVA system outperformed most of the reported results.} 
\color{black}
\begin{table*}[h!]
    \centering
    \scalebox{.7}{
    \begin{tabular}{|l|l|l|l|l|l|l|}
    \hline
       References & Energy harvesting technology& Model type & Road conditioning  & Velocity (km/h) & Dissipated power per \\ 
        ~ & ~ & ~ & ~ & ~& damper (W) \\ \hline
       Guo et al. \cite{guo2016} &Rack-pinion EM$^a$ damper with MMR$^b$& 2DOF quarter car & Class C  & 97 & $\sim$29-30 \\ \hline
         Huang et al. \cite{huang2015} & Ball-screw EM damper & SDOF quarter car & Class A &  120 & $\sim$9 \\  
         ~ & ~ & ~ & Class B &  90 & $\sim$9.5 \\  
         ~ & ~ & ~ & Class C &  50 & $\sim$21 \\  
         ~ & ~ & ~ & Class D &  30 & $\sim$24 \\  \hline
         Ataei et al. \cite{ataei2017multi} & Hybrid Hydraulic-electromagnetic damper &
        2DOF quarter car&  Class C & 50 & 32\\ \hline
         Sultoni et al. \cite{sultoni2014modeling} & Linear EM damper & 2DOF quarter car & Class C & 50 & 45 \\ \hline
         Peng et al. \cite{peng2016simulation} &Hydraulically driven EM damper& 2DOF quarter bus& Class B & 30& $\sim$42 \\ 
        ~ & ~ & 7DOF full bus & Class C & 70 & $\sim$340 \\ \hline
        Yu et al. \cite{yu2015assessment} & Rack-pinion EM damper with MMR&7DOF half car & Class C  & 50 & $\sim$15 \\ \hline
         Shi et al. \cite{singh_design_2015}& Linear EM damper with active control & 7DOF full car & Class B  & 80 & $\sim$85$^c$ \\ \hline
        Li and Zuo \cite{li2017influences} &Rack-pinion EM damper with MMR& 2DOF quarter car & Class C &  36 & $\sim$25 \\ \hline
        Yin et al. \cite{yin2016design} & Rotary EM damper with active control & 2DOF quarter car & Class B & $\sim$120 & 46 \\ \hline
         Shi et al. \cite{shi2014design} &Linear EM damper with semi-active control& 2DOF quarter car & Class C&  80 & 46.57 \\ \hline
         Tarantini \cite{tarantini2015simulation} &Ball-screw EM damper & 2DOF quarter car & Class C&  36 & $\sim$26 \\ \hline
         Bao et al. \cite{bao2016researches} &Hydraulic EHSA & 2DOF quarter car & Class C&  72 & 42.5 \\ \hline
         Chen et al. \cite{chen2020mpcbased} &Rotary EM damper with inertial nonlinearity \& MPC & 2DOF quarter car & Class C&  not mentioned & 11.04 \\ \hline
        Current Work& Ball-screw-based IPVA with MPC &2DOF quarter car & Class C & 90 & $\sim$70 to $\sim$133 \\
        ~&~&~&~&~&(perfect preview)\\\hline
    \end{tabular}}
    \caption{\textcolor{black}{Survey of simulation results amongst related works in the field.}}\label{tab:Simulation_Survey}
    \footnotesize{$^a$ EM: electromagnetic, $^b$ MMR: mechanical motion rectification, $^c$ The power is an average of 98 kJ over simulation time of 1150 s.}
\end{table*}
\textcolor{black}{\subsection{Maximizing ride comfort}}
We then examine the case of maximizing the ride comfort (i.e., minimizing the sprung mass acceleration)  where we choose  $\alpha_1=1$ and $\alpha_2=0$. The results are summarized in Fig.~\ref{minacc}. It can be seen that with perfect preview and average to high levels of SNRs, both NMPC and SL-MPC can also enhance the ride comfort. \textcolor{black}{More specifically, for the case when we have the perfect preview the average RMS value for the sprung mass acceleration is decreased by 15$\%$ and 8 $\%$ for the NMPC and SL-MPC, respectively. These numbers for the cases where SNR $=20, 15, 10$ are 11$\%$ and 7$\%$, 8$\%$ and 7$\%$, 1$\%$ and 2$\%$, respectively.}
However, with the LRDE preview, both NMPC and SLMPC fail to outperform the passive IPVA. This is because the passive design is set with the maximum damping, which leads to good ride comfort. So without accurate predictions, MPC finds it difficult to outperform the passive design regarding the ride comfort.

\begin{figure}[t!]
    \centering
    \includegraphics[width=.8\linewidth]{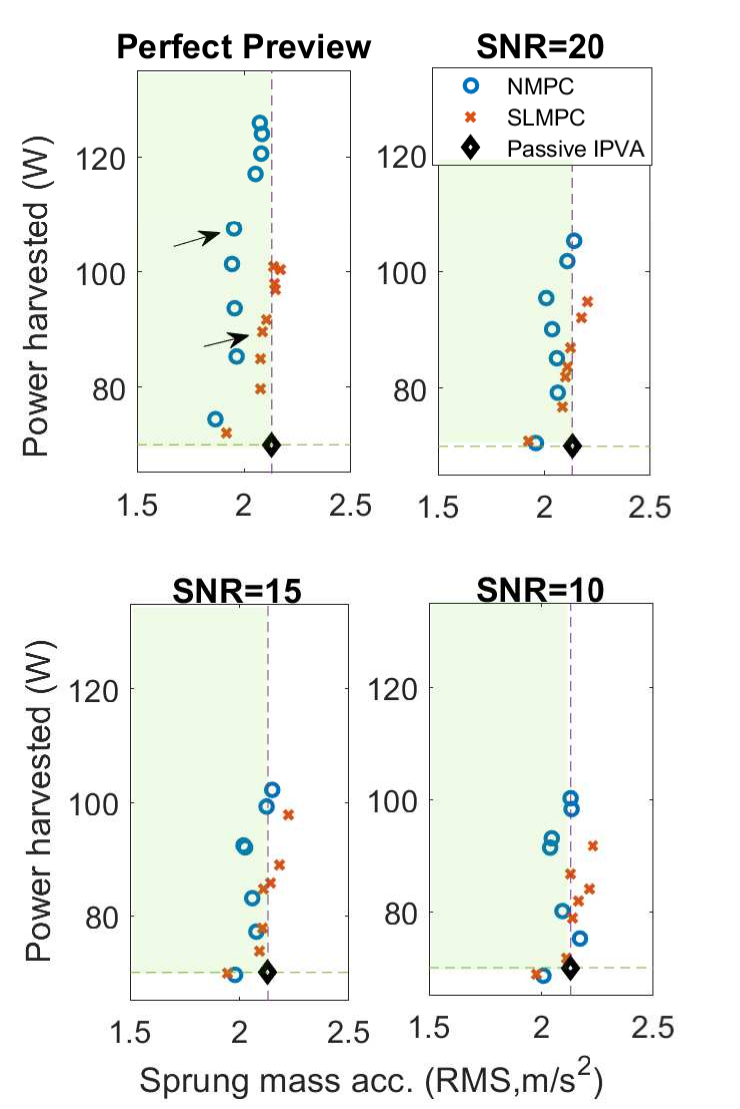}
    \caption{Control performance comparison for the case with mixed objective.}
    \label{alphas}
\end{figure}

\textcolor{black}{\subsection{Mixed objective}}
We next examine the mixed objective case, where both the power harvested and ride comfort are considered. In this case,  $\alpha_1$ is chosen as 1 and $\alpha_2$ is varied from 0.01 $\sim$ 0.1 to observe the trade-off between the power harvested and the ride comfort. Fig.~\ref{alphas} summarizes the results. For each $\alpha_2$, the SLMPC and NMPC designs are simulated 500 times and the average performance are reported. It can be seen that by varying $\alpha_2$, different trade-offs between power harvested and ride comfort can be obtained. One can choose an appropriate value that suits best for the design specifications. It can be seen that both NMPC and SL-MPC has multiple parameter settings that offer better trade-offs in  both energy harvesting and ride comfort as compared to the passive designs (i.e., in the shaded green areas). \textcolor{black}{As an example, in the perfect preview case for the circle object  pointed 
by the arrow (NMPC), the harvested power and ride comfort are improved by 55$\%$ and 9$\%$, respectively, and for the cross shaped object  pointed by the arrow (SL-MPC), the improvements are 29$\%$ and 4$\%$, respectively.} 

Lastly, we evaluate the incurred computational  complexity of the two MPC approaches. The simulations are done on a PC with a 2.5 GHz Intel Core i7-4710HQ CPU with 16GB of internal memory using MPCTools \cite{risbeck2016mpctools} and CasADi \cite{Andersson2018}. It is observed that the average computation time per 1000 steps for SL-MPC is 18 seconds, significantly less than that of NMPC, which is 43 seconds.  The results show a clear advantage of SLMPC over NMPC in terms of computation complexity,  which is a critical factor for online implementations.

\textcolor{black}{\subsection{Vehicle handling performance}}
\textcolor{black}{In this subsection, we investigate the vehicle handling performance of the proposed EHSA. For this goal the stage cost function defined in (\ref{eq:stage_cost}) is modified as follows:
\begin{equation}
     l\left( x,u \right)={{\alpha }_{1}}{{\left( {{{\dot{x}}}_{6}}+R{{{\dot{x}}}_{2}} \right)}^{2}}-{{\alpha }_{2}}u{{\left( {{x}_{2}}-{{x}_{4}} \right)}^{2}} + \alpha_3(x_5 - w)^2.
     \label{stage_cost_2}
 \end{equation}
 \begin{figure}[t!]
    \centering
    \includegraphics[width=1\linewidth]{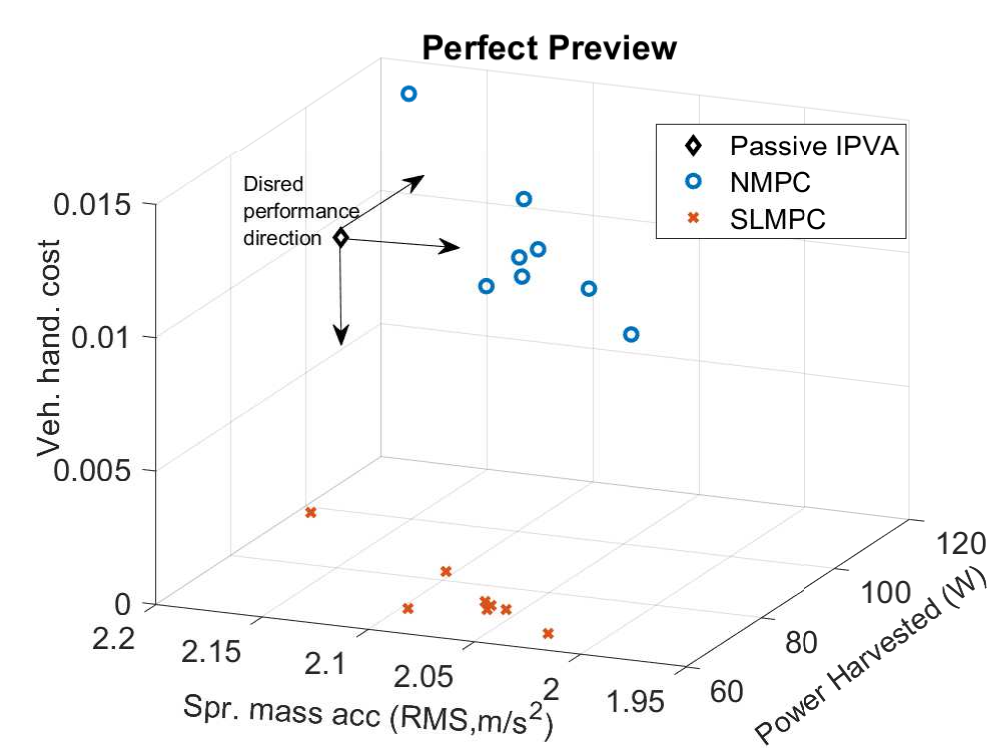}
    \caption{\textcolor{black}{Control performance comparison for the case with mixed objective.}}
    \label{fig:veh_handl}
\end{figure}
In the above equation the last term is added to consider  vehicle handling  in the control task and $\alpha_3$ is introduced to obtain different trade-offs between ride comfort, harvested energy and vehicle handling. In this regard, we considere the case which we have the perfect preview in the prediction horizon and set $\alpha_1 = 1$, $\alpha_3= 10^6$ while varyng $\alpha_2$ values in the range of  $\alpha_2 = 0.1 \sim 0.3$. The average results for 500 simulations are shown in Figure~\ref{fig:veh_handl}. It can be seen that compared with the passive IPVA, the proposed SL-MPC along with the NMPC can improve the vehicle handling metric in conjunction with metrics in energy harvesting and ride comfort. }

\section{Conclusions}
In this paper, a nonlinear IPVA was integrated into a quarter-car model, and the system with optimized parameters was shown to present simultaneous improvement in vibration control and energy harvesting. To further improve the performance, we investigated two MPC designs: nonlinear MPC (NMPC) and stochastic linearization MPC (SL-MPC), with different settings of road preview. A new SL approach was developed to stochastically  linearize systems with guaranteed stabilizability. Extensive simulations were performed which showed that SL-MPC had no major performance degradation and it significantly improved the computational efficiency.  \textcolor{black}{Specifically, our simulation results show that the power harvested has the potential to be increased by 60$\%$ and the RMS value of the sprung mass acceleration can be reduced by 7$\%$ w.r.t the passive case}. \textcolor{black}{We showed that road preview has a great impact on control performance. Future research will focus on investigating the stability of SL-MPC as well as developing a prototype system to demonstrate the proposed framework.}
\appendix

  \subsection{Linear system analytical solution}
  \label{Appendix1}
In the subsection, we derive the closed-form solution for the average power and acceleration associated with the linear benchmark shown in Fig.~\ref{fig:nonvslin}. Towards that end, Eqs.~\ref{lb1} and \ref{lb2} are first transformed into the frequency domain and the ratio of output velocity to input road velocity is determined. Namely, 
\begin{equation}
    \frac{\dot{\theta}(\omega)}{\dot{x}_r(\omega)}=
    \frac{R \omega ^2 k_t M_s}{A+Bi},
\end{equation}
where
  \begin{align*}
  A&=  
  J_r\omega^2\left[\omega^2\left(M_{us}+M_{s}\right)-k_t\right]+R^2k_s k_t\\
  &-R^2\omega ^2\left(k_t M_s+k_s M_s+k_s M_{us}-\omega^2 M_s M_{us}\right),
  \end{align*}
and
$
B=
\omega  \left[k_t-\omega ^2 \left(M_s+M_{us}\right)\right]\left(c_e+R^2 c_m\right)
$
with $i=\sqrt{-1}$.

The road disturbance $x_r$ is modeled as white noise passing through a first-order filter\cite{zuo2013energy} and the corresponding power spectral density, $S_{x_r}(\omega)$, can be written as:
\begin{equation}
    S_{x_r}(\omega)=\frac{2\pi G_r V}{\omega^2+\omega_c^2},\label{eq:psd}
\end{equation}
where $G_r$ is a road roughness coefficient specific to the class-C road as given by \cite{zuo2013energy}, $V$ is the driving speed in $m/s$ and $\omega_c$ is the cutoff frequency that keeps the power spectral density bounded at $\omega=0$. Note that $\omega_c$ needs to be  significantly smaller than the resonant frequencies of the suspension \cite{liu2016mixed} and it is assumed to be zero when deriving closed-form solutions for the linear benchmark. Otherwise, $\omega_c$ is chosen as 0.01 rad/sec. Given $\omega_c=0$, Eq.~\ref{eq:psd} implies that the time derivative of the road disturbance (i.e., $\dot{x}_r$) is a white noise with intensity of $2\pi G_rV$.
Noting that the instantaneous power is given by $P=c_e\dot{\theta}^2$ and proceeding with the principle of $H_2$ norm \cite{roberts2003random}, the average power becomes:

\begin{equation}\label{rms power}
    P=
    \frac{c_e}{2\pi}\int_{-\infty}^\infty 2\pi G_rV\left|\frac{\dot{\theta}}{\dot{x}_r}\right|^2d\omega=\frac{c_e\pi  V G_r k_t}{c_e+R^2 c_m}.
\end{equation}
Likewise, the RMS sprung mass acceleration, $\sigma_{\Ddot{x}_s}$, can be found by first deriving
\begin{eqnarray}
\frac{\ddot{x}_s(\omega)}{\dot{x}_r(\omega)}
=
   k_t\omega\frac{\left[i \left(k_s R^2-\omega ^2 J_r\right)-\omega\left(c_e+c_mR^2\right)\right]}{A+Bi},
\end{eqnarray}
where A and B were previously defined.  \color{black}The mean square acceleration can then be determined by integrating the power spectral density for the acceleration over the entire frequency domain, noting that the power spectral density value for $\dot{x}_r$ is $ S_{\dot{x}_r}(\omega)=2\pi G_rV$ and the power spectral density of the sprung mass acceleration is $ S_{\ddot{x}_s}(\omega)=\left|\frac{\ddot{x}_s}{\dot{x}_r}\right|^2S_{\dot{x}_r}(\omega)$. \color{black}The RMS value is then simply defined as
\begin{eqnarray}\label{rms acc}
    \sigma_{\Ddot{x}_s}=\sqrt{\frac{1}{2\pi}\int_{-\infty}^\infty S_{\ddot{x}_s}(\omega)d\omega}
    =\sqrt{\frac{1}{2\pi}\int_{-\infty}^\infty 2\pi G_rV\left|\frac{\ddot{x}_s}{\dot{x}_r}\right|^2d\omega}.
\end{eqnarray}
Eqn.~\ref{rms acc} can then be integrated with the known integration formula\cite{roberts2003random} to obtain

\begin{equation}\label{accelsol}
\begin{aligned}
\sigma_{\Ddot{x}_s}&=\sqrt{\pi VG_r C^{-1}\left(a_0+a_1R^2+a_2R^4+a_3R^6\right)},
\end{aligned}
\end{equation}
where $M=M_s+M_{us}$ corresponds to the total mass and 
\[C=R^2 M_s^2 \left(c_e+R^2 c_m\right) \left(J_r M+R^2 M_s M_{us}\right),\]
\[a_0=J_rk_tc_e^2 M+J_r^3 k_t^2,\]
$$a_1=2 c_e c_m J_r Mk_t+c_e^2 M_s M_{us}+k_t^2 M_s-J_r^2 2 k_s M k_t,$$
$$a_2=2 c_e c_m k_t M_s M_{us}+c_m^2 J_r k_t M+k_s^2 J_r M^2
-2 k_tk_s J_r M_s M_{us},$$ 
$$a_3=M_s M_{us}c_m^2 k_t+k_s^2M_s M_{us} M.$$
Equations \ref{rms power} and \ref{accelsol} are then used to plot the closed form solution shown in Fig. \ref{fig:nonvslin}.

\subsection{ Stochastic linearization with guaranteed stability}
In this subsection we advance the SL approach so that the linearized model (e.g., Eqn.~\ref{eq:LTI}) has guaranteed stabilizability.
 More specifically, we aim at reformulating the problem to find a matrix $A_{ls}$ such that  $(A_{ls},B_l)$ is stabilizable and is close to $A_l$ (e.g., $A_l=A_{ls}$ if $(A_{l},B_l)$ is already stabilizable). We next review some definitions regarding stabilizability  to place our method in proper context.
 
 \noindent
 \textit{Definition 1.} Consider a linear time invariant (LTI) system $\dot{x}=Ax+Bu$. The pair $(A,B)$ is called stabilizable if all its uncontrollable eigenvalues are stable. 

The following lemma can be used to check the stabilizability of an LTI system.

\noindent
\textit{Lemma 1 \cite{antsaklis2006linear}.}  Consider an LTI system $\dot{x}=Ax+Bu,$ 
where $A$ is $n\times n$ and $B$ is $n\times m$. The controllability matrix $\mathcal{C}$ is defined as $\mathcal{C} = [B, AB,\dots A^{n-1}B]$.
If rank $\mathcal{C}  = q < n$, then there exists a nonsingular $n\times n$ matrix $T$
such that 
\[\hat{A}={{T}^{-1}}AT=\left[ \begin{matrix}
   {{{\hat{A}}}_{11}} & {{{\hat{A}}}_{12}}  \\
   0 & {{{\hat{A}}}_{22}}  \\
\end{matrix} \right],\text{   }\hat{B}={{T}^{-1}}B=\left[ \begin{aligned}
  & {{{\hat{B}}}_{1}} \\ 
 & 0 \\ 
\end{aligned} \right].\]
where $\hat{A}_{11}$ is $q\times q$, $\hat{B}_{1}$ is $q\times m$, and the pair $(\hat{A}_{11},\hat{B}_{1})$ is controllable. The transformation matrix $T$ is called controllability decomposition matrix and can be constructed as follows. Let $T=\left[ \begin{matrix}
  X & Y \end{matrix} \right] $, then $X$ is is an $n\times q $ matrix whose columns span the
columns space of $\mathcal{C}$,  and $Y$ is an $n\times(n-q)$ matrix whose columns are chosen such that $T$ is nonsingular.

The eigenvalues of $\hat{A}_{11}$ are called the controllable eigenvalues and
those of $\hat{A}_{22}$ are called the uncontrollable eigenvalues. Following Lemma 1, if we transform ${A}_{l}$  as $\hat{A}_{l}={{T}^{-1}}A_{l}T=\left[ \begin{matrix}
   {{{\hat{A}}}_{l,11}} & {{{\hat{A}}}_{l,12}}  \\
   0 & {{{\hat{A}}}_{l,22}} \\
   \end{matrix} \right]$,
then ${A}_{l}$ is stabilizable if and only if $\hat{A}_{l,22}$ is stable. If this is not the case, then we seek a Hurwitz matrix $\hat{A}_{ls,22}$ that is similar to $\hat{A}_{l,22}$ (e.g., in the sense of matrix norms) to replace $\hat{A}_{l,22}$. 
If $\hat{A}_{l,22}$ is not Hurwitz, we aim for finding the Hurwitz matrix $A_{eq}$ where its second norm difference with $\hat{A}_{l,22}$ is minimized. To this end, using the Lyapunov inequality for verifying the stability of a linear system \cite{khalil2002nonlinear}, the optimization problem is such that
\begin{equation}
\label{eq:optbmi}
\begin{matrix}
  \underset{{{A}_{eq}},P}{\mathop{\min.}}\,\text{   }\left\| {{A}_{eq}}-{{{\hat{A}}}_{l,22}} \right\|_{2}^{2}\text{ } \\ 
  \text{s}\text{.t}\text{.  }A_{eq}^{T}P+P{{A}_{eq}}<0,\text{ }P>0. 
\end{matrix}
\end{equation}

Two decision variables are multiplied together in the first constraint, making the problem a BMI (bilinear matrix inequality) optimization problem, which can be solved using the MATLAB BMIsolver package. After finding $A_{eq}$, this matrix will replace $\hat{A}_{l,22}$ in $\hat{A}_{l}$ and one can use the matrix $T$ to transform back and get the matrix ${A}_{l}$ where the pair (${A}_{l},B_{l}$) is now stabilizable.

\subsection{High gain disturbance observer design }
In this subsection, we show the development of high-gain observer (HGO \cite{khalil2014high}) to estimate the road disturbance that is exploited in the above MPC formulations in the prediction horizon.  Specifically, we rewrite the dynamics of the controlled energy harvesting system in a compact form based on Eq. \ref{Eq:kavaguchi} as follows:
\begin{equation}
    \begin{aligned}
  & {{{\dot{x}}}_{1}}={{x}_{2}}, \text{ } {{{\dot{x}}}_{2}}={{f}_{2}}(x,u,w ), \\ 
 & {{{\dot{x}}}_{3}}={{x}_{4}}, \text{ } {{{\dot{x}}}_{4}}={{f}_{4}}(x,u,w ), \\ 
 & {{{\dot{x}}}_{5}}={{x}_{6}}, \text{ } {{{\dot{x}}}_{6}}={{f}_{6}}(x,u,w). \\ 
\end{aligned}
\label{eq:nrmfrm}
\end{equation}
where $f_{6}(x,u,w)={{b}_{1}}(x,u)+w{{b}_{2}}({x})$.

\begin{figure}[!h]
    \centering
    \includegraphics[width=.9\linewidth]{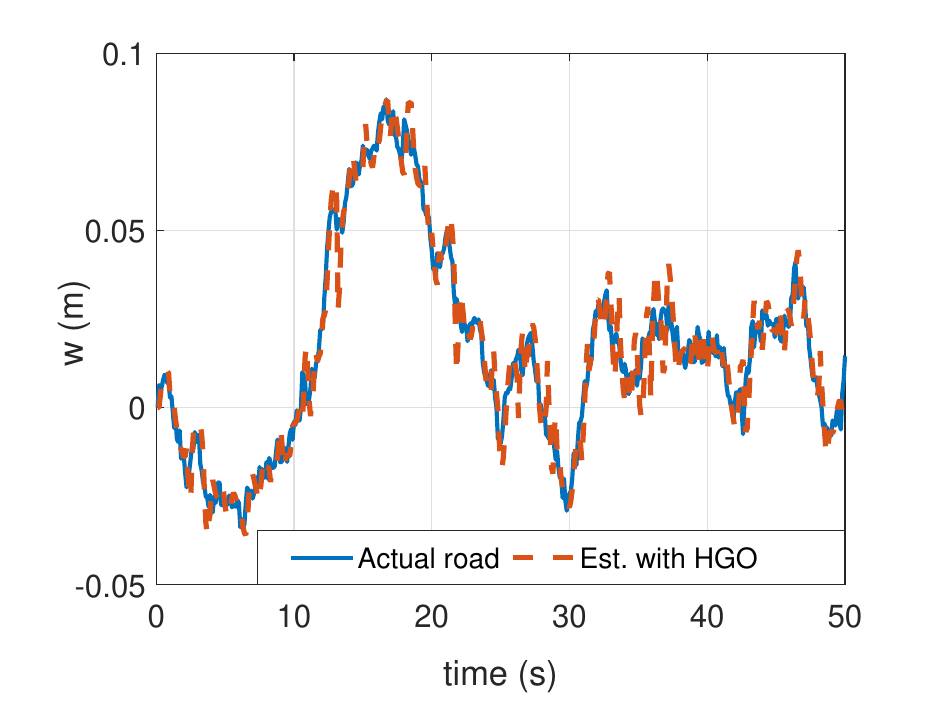}
    \caption{Road profile estimation using HGO}
    \label{fig:rHGO}
\end{figure}

Eqn.~\ref{eq:nrmfrm} consists of 3 separate sub-dynamics in the normal nonlinear SISO form with $y_1=x_1$, $y_2=x_3$ and $y_3=x_5$ being our measurements. In this regard, in order to design the extended observer, we take $\sigma=f_6(x,u,w )$ as the function to be estimated. Thus, the observer equations are formulated as follows:

\begin{equation}
    \begin{aligned}
  & {{{\dot{\hat{x}}}}_{1}}={{{\hat{x}}}_{2}}+\left( {{\alpha }_{1}}/{{\varepsilon }_{1}} \right)\left( {{y}_{1}}-{{{\hat{x}}}_{1}} \right), \\ 
 & {{{\dot{\hat{x}}}}_{2}}=\left( {{\alpha }_{2}}/{{\varepsilon }_{1}}^{2} \right)\left( {{y}_{1}}-{{{\hat{x}}}_{1}} \right), \\ 
 & {{{\dot{\hat{x}}}}_{3}}={{{\hat{x}}}_{4}}+\left( {{\alpha }_{3}}/{{\varepsilon }_{2}} \right)\left( {{y}_{2}}-{{{\hat{x}}}_{3}} \right) \\ 
 & {{{\dot{\hat{x}}}}_{4}}=\left( {{\alpha }_{4}}/\varepsilon _{2}^{2} \right)\left( {{y}_{2}}-{{{\hat{x}}}_{3}} \right), \\ 
 & {{{\dot{\hat{x}}}}_{5}}={{{\hat{x}}}_{6}}+\left( {{\alpha }_{5}}/{{\varepsilon }_{3}} \right)\left( {{y}_{3}}-{{{\hat{x}}}_{5}} \right), \\ 
 & {{{\dot{\hat{x}}}}_{6}}={{{\hat{\sigma}}}}+\left( {{\alpha }_{6}}/{{\varepsilon }_{3}}^{2} \right)\left( {{y}_{3}}-{{{\hat{x}}}_{5}} \right), \\ 
 & \dot{\hat{\sigma }}=\left( {{\alpha }_{7}}/{{\varepsilon_3}}^{3} \right)\left( {{y}_{3}}-{{{\hat{x}}}_{5}} \right). \\ 
\end{aligned}
\end{equation}
Here $\varepsilon _{1}, \varepsilon_{2}$ and $\varepsilon_{3}$ are sufficiently small positive constants;  $\alpha_{i}$ and $\alpha_{i+1}$, $i=1,2$, are chosen such that $s^2+\alpha_{i}s+\alpha_{i+1}$ is Hurwitz; and $\alpha_{5}$, $\alpha_{6}$ and $\alpha_{7}$ are also chosen such that $s^3+\alpha_{5}s^2+\alpha_{6}s+\alpha_{7}$ is Hurwitz. 
Then at time $t$, the disturbance can be estimated as \cite{khalil2014high}:
$$   \hat{w}(t)= \frac{\hat{\sigma}(t)-{b_{1}(\hat{x}(t),u(t))}}{{{b}_{2}}({{\hat{x}(t) }})}.$$ A sample of the estimation is shown in Fig. \ref{fig:rHGO}.

\bibliographystyle{ieeetr}
\bibliography{main}

\begin{IEEEbiography}
 [{\includegraphics[width=1in,height=1.25in,clip,keepaspectratio]{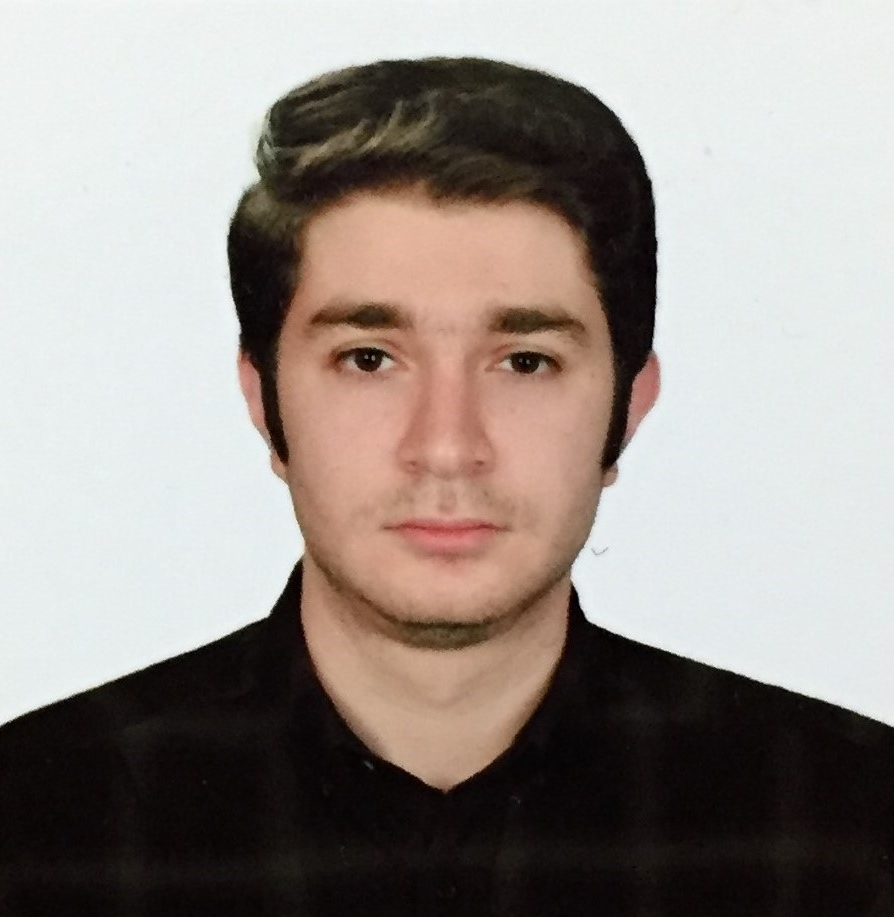}}]{Mohammad R. Hajidavalloo} obtained his B.Sc. and M.Sc. degree from  University of Tehran in Mechanical Engineering  in 2016 and 2018 respectively.

He is currently pursuing the Ph.D. degree in the department of Mechanical Engineering at Michigan State University. His research interests include Learning-based Control, Optimal Control and Automated Vehicles.
\end{IEEEbiography}

\begin{IEEEbiography}
 [{\includegraphics[width=1in,height=1.25in,clip,keepaspectratio]{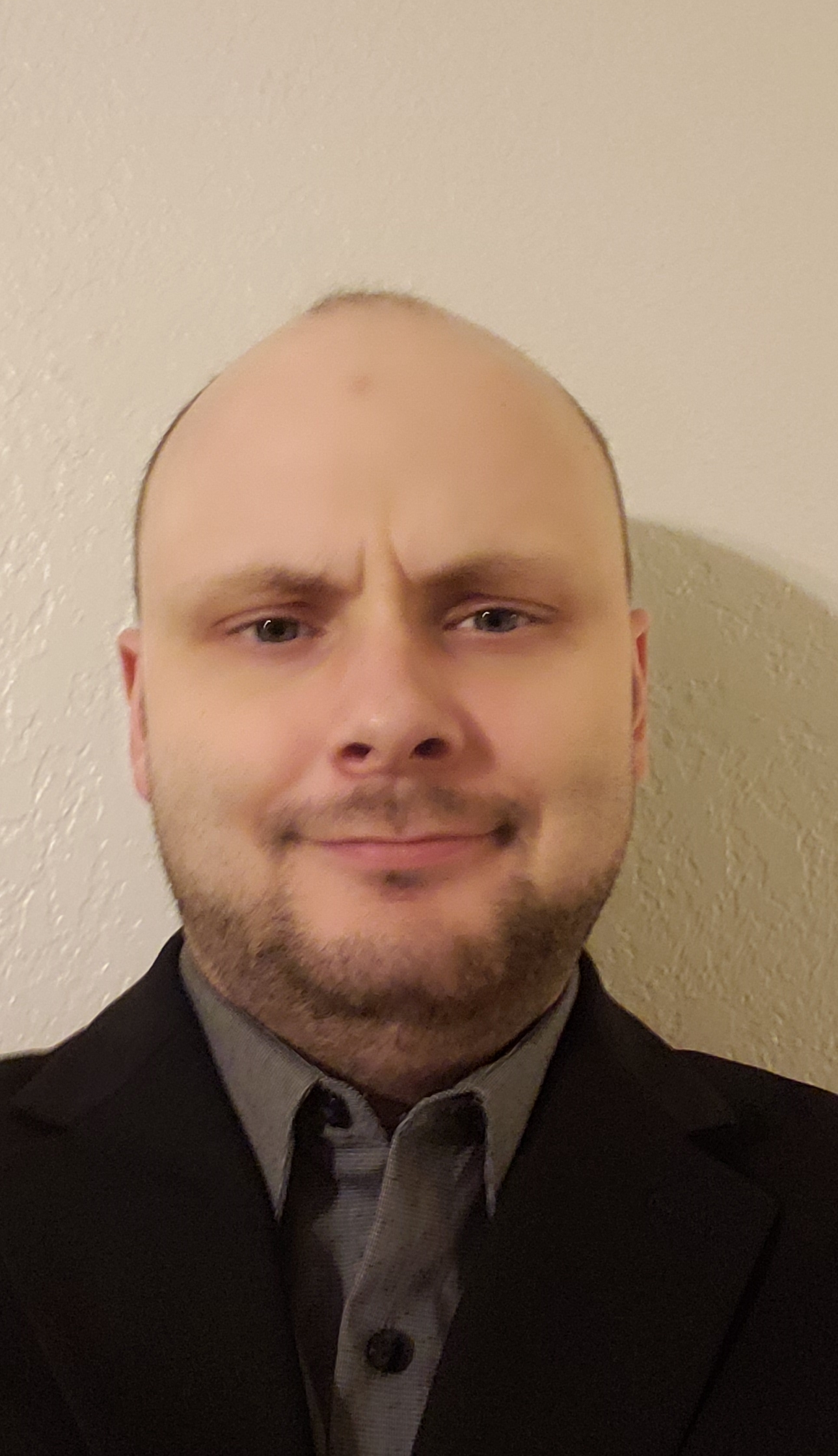}}]{Joel Cosner} received his B.S. degree from Michigan State University in 2013.
 
 He is currently pursuing a Ph.D. in Mechanical Engineering at Michigan State University. His research interests include Nonlinear Dynamics, Vibration Suppression, and Energy Harvesting. 
\end{IEEEbiography}

\begin{IEEEbiography}
 [{\includegraphics[width=1in,height=1.25in,clip,keepaspectratio]{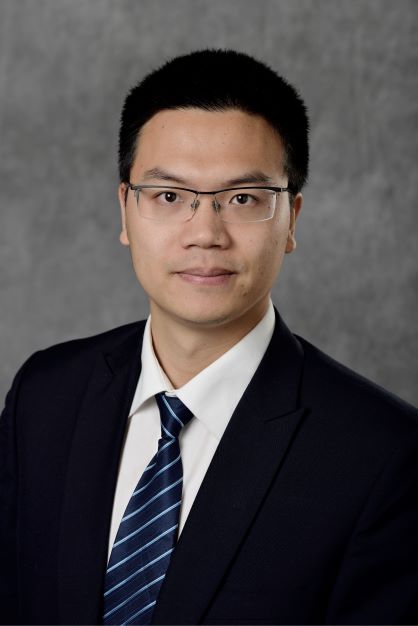}}]{Zhaojian Li} received his B. Eng. degree from Nanjing University of Aeronautics and Astronautics in 2010. He obtained M.S. (2013) and Ph.D. (2015) in Aerospace Engineering (flight dynamics and control) at the University of Michigan, Ann Arbor.

He is currently an Assistant Professor with the department of Mechanical Engineering at Michigan State University. His research interests include Learning-based Control, Nonlinear and Complex Systems, and Robotics and Automated Vehicles. He is a senior member of IEEE and a recipient of the NSF CAREER Award.
\end{IEEEbiography}

\begin{IEEEbiography}
 [{\includegraphics[width=1in,height=1.25in,clip,keepaspectratio]{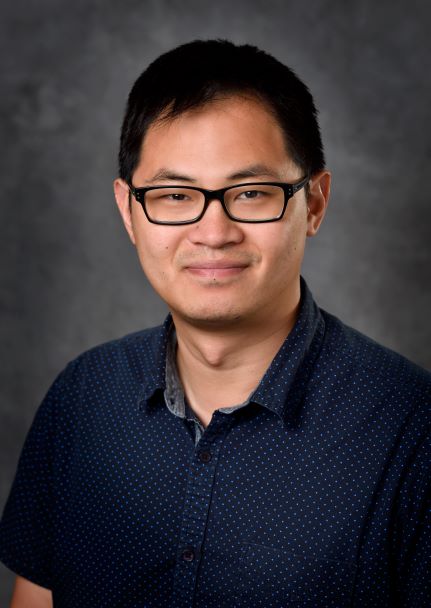}}]{Wei-Che Tai} received his B.S. degree from National Taiwan University in 2007. He obtained M.S. (2012) and Ph.D. (2014) in Mechanical Engineering at the University of Washington, Seattle. 
He is currently an Assistant Professor with the department of Mechanical Engineering at Michigan State University. His research interests include Nonlinear and Stochastic Vibration, Rotordynamics, and Energy Harvesting. He has served on the ASME Technical Committee on Vibration and Sound since 2019 and ASME Technical Committee on Energy Harvesting since 2020.
\end{IEEEbiography}
\vspace{-300pt}
\begin{IEEEbiography}
 [{\includegraphics[width=1in,height=1.25in,clip,keepaspectratio]{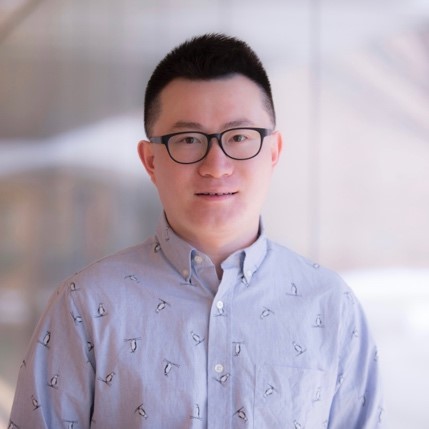}}]{Ziyou Song} is an Assistant Professor with the Department of Mechanical Engineering at the National University of Singapore (NUS). He received B.E. degree (with honours) and Ph.D. degree (with highest honours) in Automotive Engineering from Tsinghua University, Beijing, China, in 2011 and 2016, respectively. After graduation, he worked as a Research Scientist at Tsinghua University from 2016-2017. From 2017 to 2019, he worked as a Postdoctoral Research Fellow at the University of Michigan, Ann Arbor, where he was also an Assistant Research Scientist/Lecturer from 2019 to 2020. Prior to joining NUS, he was a Battery Algorithm Engineer at Apple Inc., Cupertino, US. 
Dr Song’s research interests lie in the areas of modelling, estimation, optimization, and control of energy storage (e.g., battery, supercapacitor, and flywheel) for electrified vehicles and renewable energy systems. He is the author or co-author of 2 book chapters and more than 60 peer-reviewed publications. He has received several paper awards, including Applied Energy 2015-2016 Highly Cited Paper Award, Applied Energy Award for Most Cited Energy Article from China, NSK Outstanding Paper Award of Mechanical Engineering, and 2013 IEEE VPPC Best Student Paper Award. Dr. Song serves as reviewer for more than 40 international journals. He also serves as Associate Editor for Automotive Innovation, SAE International Journal of Electrified Vehicles and IEEE Transactions on Transportation Electrification.
\end{IEEEbiography}

\end{document}